\documentclass[aps,prl,doublecolumn,superscriptaddress,a4paper,reprint]{revtex4-1}
\pdfoutput=1
\usepackage{ifpdf}
\usepackage[latin1]{inputenc}
\usepackage{graphicx}
\usepackage{dcolumn}
\usepackage{bm}
\usepackage{amsmath}
 
\graphicspath{{Figures/}}
\usepackage[dvipsnames]{xcolor}
\usepackage{cancel}
\usepackage{braket}
\colorlet{ForestGreen}{black}
\colorlet{Blue}{black}
\colorlet{BurntOrange}{black}
\colorlet{Sepia}{black}

\begin{document}


\title{Magnon-exciton proximity coupling at a van der Waals heterointerface}
\author{A. Gloppe}
\email{arnaud.gloppe@ipcms.unistra.fr}
\affiliation{Universit\'e de Strasbourg, CNRS, Institut de Physique et Chimie des Mat\'eriaux de Strasbourg (IPCMS), UMR 7504, F-67000 Strasbourg, France}
\affiliation{Research Center for Advanced Science and Technology (RCAST), The University of Tokyo, Meguro-ku, Tokyo 153-8904, Japan}
\author{M. Onga}
\affiliation{Quantum-Phase Electronics Center (QPEC) and Department of Applied Physics, The University of Tokyo, Tokyo 113-8656, Japan}
\author{R. Hisatomi}
\affiliation{Research Center for Advanced Science and Technology (RCAST), The University of Tokyo, Meguro-ku, Tokyo 153-8904, Japan}

\author{A. Imamo\u{g}lu}
\affiliation{Institute for Quantum Electronics, ETH Zurich, CH-8093, Zurich, Switzerland}

\author{Y. Nakamura}
\affiliation{Research Center for Advanced Science and Technology (RCAST), The University of Tokyo, Meguro-ku, Tokyo 153-8904, Japan}
\affiliation{RIKEN Center for Quantum Computing (RQC), Wako, Saitama 351-0198, Japan}

\author{Y. Iwasa}
\email{iwasa@ap.t.u-tokyo.ac.jp}
\affiliation{Quantum-Phase Electronics Center (QPEC) and Department of Applied Physics, The University of Tokyo, Tokyo 113-8656, Japan}
\affiliation{RIKEN Center for Emergent Matter Science (CEMS), Wako, Saitama 351-0198, Japan}

\author{K. Usami}
\email{usami@qc.rcast.u-tokyo.ac.jp}
\affiliation{Research Center for Advanced Science and Technology (RCAST), The University of Tokyo, Meguro-ku, Tokyo 153-8904, Japan}

\date{\today}

\begin{abstract}
Spin and photonic systems are at the heart of modern information devices and emerging quantum technologies~\cite{Otani2017,Hammerer2010,Clerk2020}. An interplay between electron-hole pairs (excitons) in semiconductors and collective spin excitations (magnons) in magnetic crystals would bridge these heterogeneous systems, leveraging their individual assets in novel interconnected devices. Here, we report the magnon--exciton coupling at the interface between a magnetic thin film and an atomically-thin semiconductor. Our approach allies the long-lived magnons hosted in a film of yttrium iron garnet (YIG)~\cite{Kajiwara2010, Tabuchi2015} to strongly-bound excitons in a flake of a transition metal dichalcogenide, MoSe$_2$~\cite{Xiao2012,Wang2018,Mak2018,Mak2016}. The magnons induce on the excitons a dynamical valley Zeeman effect ruled by interfacial exchange interactions. This nascent class of hybrid system suggests new opportunities for information transduction between microwave and optical regions. 
\end{abstract}

\maketitle
\paragraph*{}
The emergence of 2D materials, such as graphene, sheds a new light on condensed-matter physics, for these materials can be artificially stacked to combine, protect or enhance individual pristine physical properties~\cite{Geim2013}. 
Atomically-thin semiconducting transition metal dichalcogenides (TMDs), exhibiting a number of unique optical features such as large excitonic binding energies and valley-contrasting exciton selection rules~\cite{Xiao2012,Wang2018,Mak2018}, attract a lot of attention as a new platform for quantum optics and nanophotonics~\cite{Mak2016,Lundt2019, Lorchat2020}. By a static valley Zeeman effect, their exciton resonances are shifted by external magnetic fields~\cite{Li2014,Aivazian2015,Srivastava2015} or by the interfacial exchange fields with a magnetic substrate~\cite{Zhao2017,Norden2019,Qi2015,Scharf2017}. 

In this work, we study a heterostructure consisting of MoSe$_2$ flakes transferred on a magnetic film made of yttrium iron garnet (Fig.~1\textbf{a}). The film supports long-lived magnons, or magnetization oscillations~\cite{Stancil2009}, that can be coherently driven by microwaves. 
Magnons play a major role in spintronics circuits as a low-loss information carrier~\cite{Kajiwara2010,Chumak2015}, and in quantum hybrid systems as a macroscopic quantum interface to superconducting quantum bits~\cite{Tabuchi2015,Clerk2020,Lachance-Quirion2020}.
\color{ForestGreen}
Realizing an interfacial coupling between magnons and excitons, in contrast to previous efforts involving bulk magnets~\cite{Freeman1968,Meltzer1968} or dilute ferromagnetic semiconductors~\cite{Nemec2012,Dietl2014}, offers a promising way forward to connect these technologies to optics.
\color{ForestGreen}Here, we demonstrate this magnon--exciton coupling at a van der Waals heterointerface. \color{black} 

\begin{figure}
\includegraphics[scale=1]{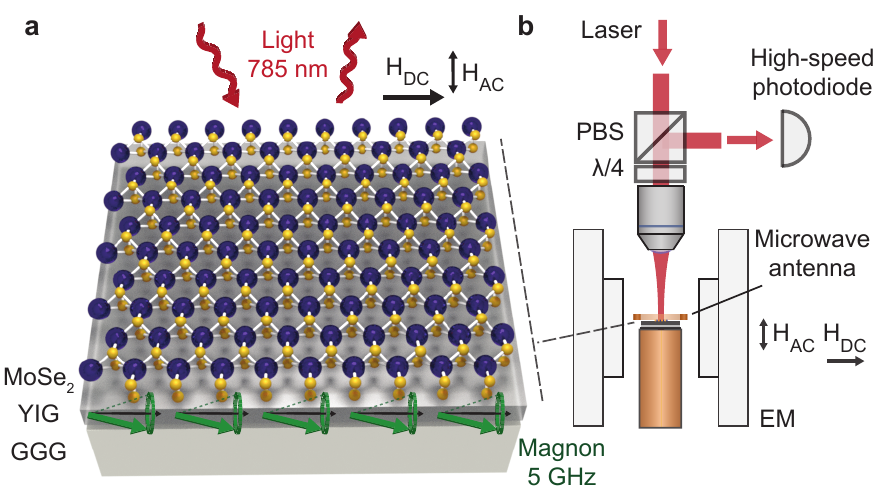}
\caption{\label{figure1} \textbf{Optically-addressed TMD flake on a magnetic substrate supporting magnon modes.} 
\textbf{a,} Atomically-thin flakes of MoSe$_2$ are stacked on a magnetic YIG film grown on gadolinium gallium garnet (GGG, see Methods). The magnetization of the film is saturated by a static magnetic field $H_\mathrm{DC}$ directed within its plane. A microwave antenna excites magnons of the fundamental magnetostatic mode of frequency $\omega_m/2\pi\sim~5\,$GHz through the alternating magnetic field $H_\mathrm{AC}$. \textbf{b,} The flakes are addressed normally with a focused laser beam at $\lambda_L = 785\,$nm with a left- or right-handed circular polarization. A high-speed photodetector detects the optical signal reflected off the sample with the same polarization as input (EM: electromagnet, PBS: polarizing beamsplitter, $\lambda/4$: quarter-wave plate, see Methods).}
\end{figure}

\begin{figure}
\includegraphics[scale=1]{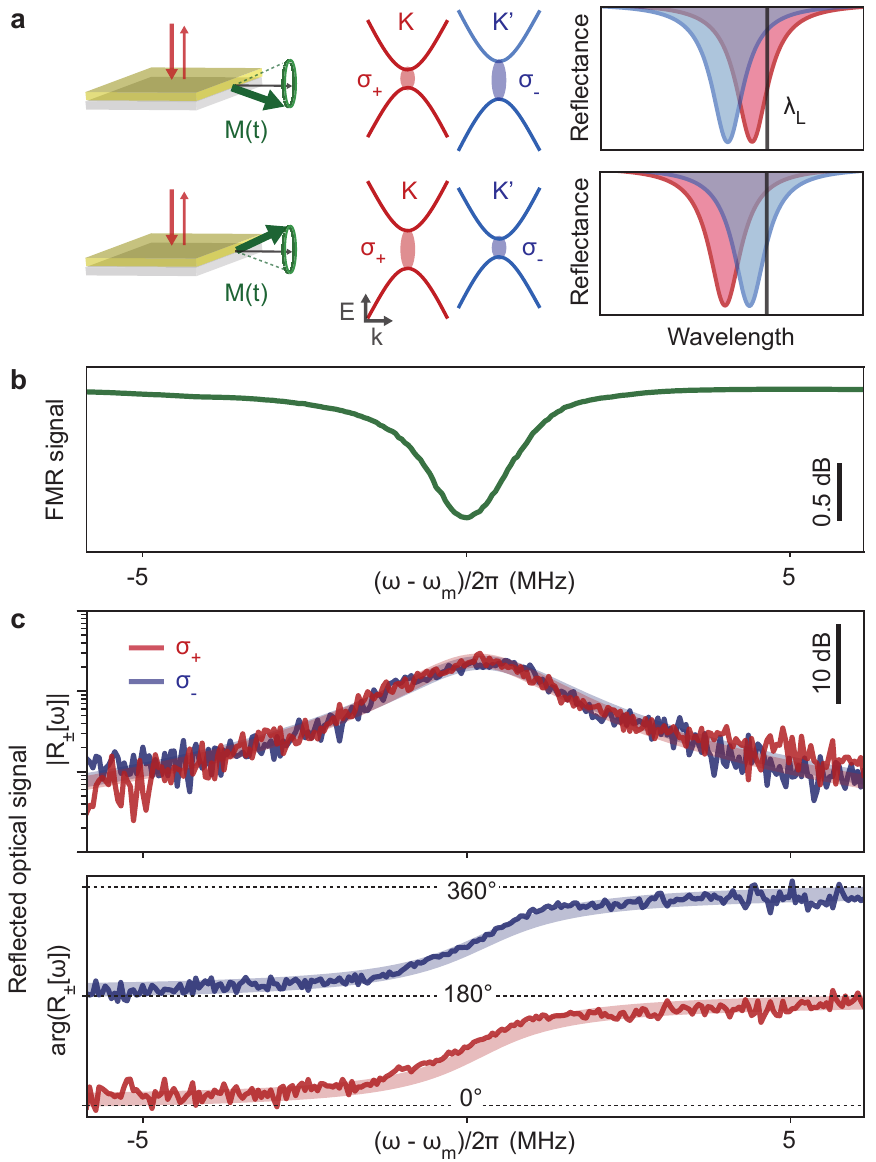}
\caption{\label{figure2} \textbf{Dynamical valley Zeeman effect.} 
\textbf{a,} 
Magnons support a coherent oscillation of the magnetization vector $M(t)$, responsible for an effective magnetic field modulating the excitonic resonances of the TMD flake through a dynamical valley Zeeman effect. 
Carrying opposite magnetic moments, the two valleys K and K$'$ have their excitonic resonance shifting opposite ways when experiencing an out-of-plane magnetic field ($E$: relative energy, $k$: electron momentum).  
The \textcolor{black}{reflectance} of the flake at a fixed laser wavelength $\lambda_L$, on the edge of the excitonic resonance, is subsequently modulated at the magnon frequency. The phase of the reflected signal depends on the valley index, selectively addressed with left-handed $\sigma_+$ and right-handed $\sigma_-$  circularly polarized light for K and K$'$ valleys, respectively. 
\textbf{b,}~Microwave absorption signal revealing the ferromagnetic resonance (FMR) at $\omega_m/2\pi = 5.64$\,GHz. 
\textbf{c,} Magnitude and relative phase of the optically-probed FMR spectra~$R_\pm[\omega]$ on a MoSe$_2$ flake.  
The spectrum with the left(right)-handed circularly polarized light is plotted as a solid red (blue) line, superimposed on a red (blue) translucent Lorentzian fit (flake thickness: 20\,nm, $n_\mathrm{magnon} = 10^{14}$, detection bandwidth: 5\,Hz). 
}
\end{figure}

\paragraph*{}
Figure~\ref{figure1}\textbf{b} depicts schematically our experimental setup at room temperature. The heterostructure MoSe$_2$/YIG is placed in the gap of an electromagnet to saturate the YIG magnetization within the film plane ($H_\mathrm{DC} \sim 0.1\,$T$/\mu_0$). Magnons in the uniform magnetostatic mode of the film are excited with a microwave loop-antenna connected to a network analyzer. A typical spectrum of the microwave reflected off the antenna  reveals the ferromagnetic resonance (FMR) in Fig.~\ref{figure2}\textbf{b}, centered at $\omega_m/2\pi = 5.64$\,GHz with a linewidth $\gamma_m/2\pi= 1.75$\,MHz. With a continuous-wave laser ($\lambda_L = 785$\,nm) we examine the light reflected from the heterostructure on a high-speed photodiode.

\paragraph*{}
By analogy to valley Zeeman effects observed in TMDs with a static magnetic field~\cite{Li2014,Aivazian2015,Srivastava2015,Zhao2017,Norden2019}, we expect the out-of-plane magnetization oscillations due to driven magnons to shift dynamically the excitonic resonance. \textcolor{ForestGreen}{The transverse magnetization per single magnon depends on the magnetic film volume, as our magnetostatic mode is delocalized over the entire magnetic film (see SI).}
The relevant interaction Hamiltonian can be written as 
\begin{equation}
H = \tau  \, \hbar g \left(\frac{\hat{a} + \hat{a}^\dagger}{2} \right) \hat{x}^\dagger\hat{x}
\end{equation}
where $\tau = \pm 1$ is the index for K and K$'$ valleys, $\hat{a}$ ($\hat{a}^\dagger$) and $\hat{x}$ ($\hat{x}^\dagger$) are the annihilation (creation) operators for magnon and exciton, respectively  (see SI). The magnon--exciton coupling rate $g$ corresponds to the excitonic resonance shift induced by a single magnon. \textcolor{ForestGreen}{This could stem from a short-range interfacial exchange field or from a long-range dipolar field.} \textcolor{Blue}{The microscopic origin of the coupling will be discussed later.}

The choice of MoSe$_2$ is motivated by its bright excitons, with a high emission yield and a fundamental resonance around 800\,nm~\cite{Arora2015} corresponding to a low-absorption window for YIG. 
In the absence of external magnetic field, photoluminescence measurements (see SI) show that the excitonic resonances of the MoSe$_2$ flakes are not significantly affected by their stacking on the YIG and match the typical values obtained on Si/SiO$_2$ substrates~\cite{Arora2015}.
The optical \textcolor{black}{reflectance} of the flake at a fixed wavelength, situated on the edge of the excitonic resonance, should be subsequently modulated at the frequency of the magnetization oscillations (Fig.~\ref{figure2}\textbf{a}), constituting a signature of the dynamical valley Zeeman effect. \textcolor{Sepia}{The tenuous Zeeman shift induced by the magnons at microwave frequencies would be otherwise challenging to observe with conventional optical spectroscopy methods.}

\begin{figure}
\includegraphics[scale=1]{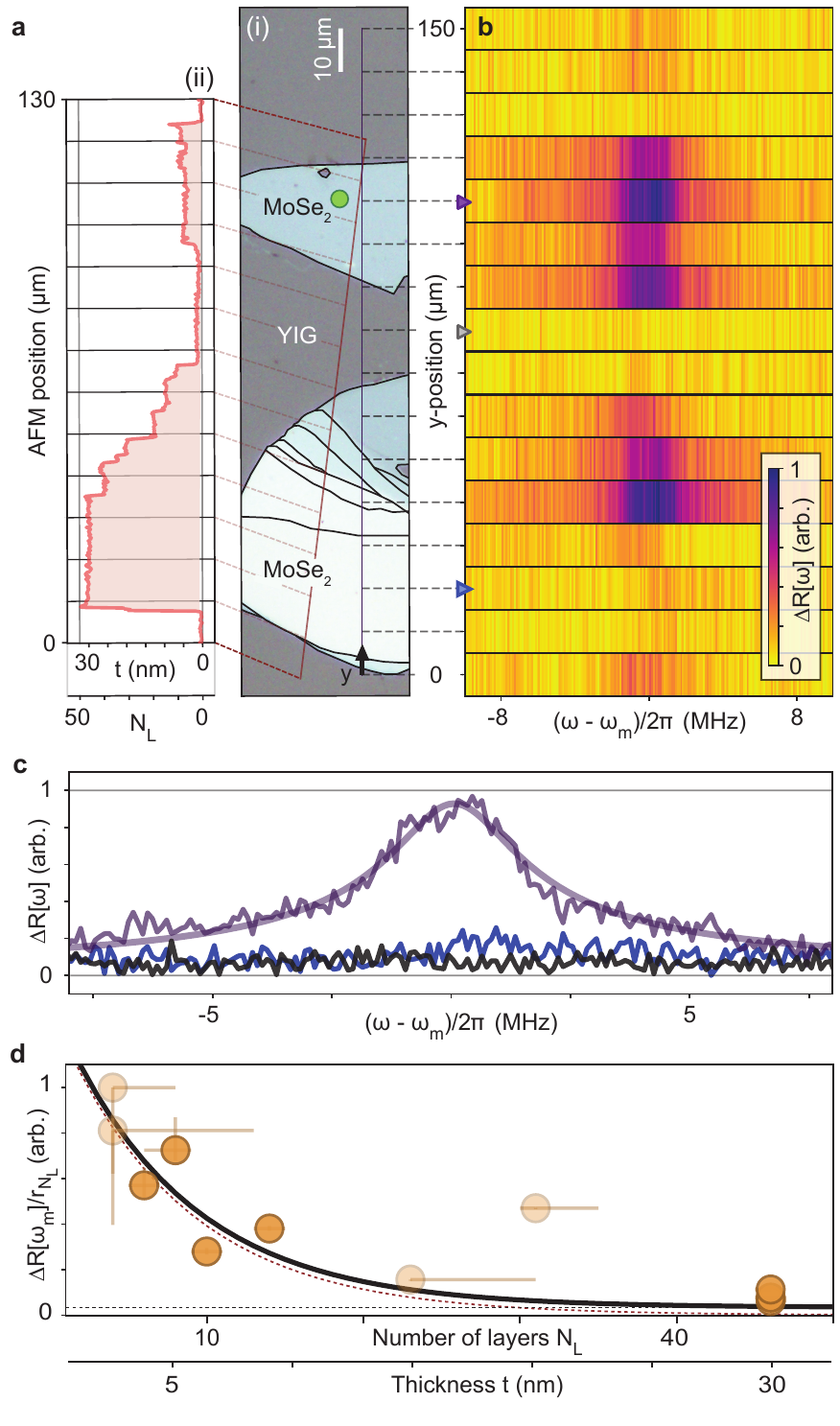}
\caption{\label{figure3} \textbf{Thickness dependence of the magneto-optical response.} 
\textbf{a,} Optical micrograph of the heterostructure under white light illumination (i) abutting a topography measurement along the dark orange line realized with an atomic force microscope (ii), where $t$ is the calibrated thickness and $N_L$ is the deduced number of MoSe$_2$ layers (see Methods). 
Black contours highlight change in optical contrast. A green circle marks the region investigated in the calibrated measurements in Fig.~4. The horizontal dashed lines along the $y$-axis mark the successive positions of the laser spot center (waist radius: $2\,\mu$m). \textbf{b,} Differential optical reflectance $\Delta R[\omega] = |R_+[\omega] - R_-[\omega]|$ around the magnon frequency as a function of the laser position on the sample. The spectra at positions $y = 80\,\mu$m (black), $y = 20\,\mu$m (blue) and $y = 110\,\mu$m (purple) are shown in \textbf{c}, corresponding respectively to bare YIG, a thick  MoSe$_2$ flake ($N_L \sim 46$) and a few-layer MoSe$_2$ flake ($N_L \sim 8$). The purple translucent line is a Lorentzian fit ($\omega_m/2\pi = 5.58$\,GHz). \textbf{d,} Decay of the normalized differential optical reflectance with the MoSe$_2$ number of layers while driving magnons at $\omega_m/2\pi$. The solid thick line is a fit with a model of the fraction of light coming back from the very bottom layer \textcolor{Blue}{(red dotted line)} and a plateau (black dotted line) (see Methods and SI). Data with large error bars in brown, marked as translucent, are not used for the fit (see Methods).}
\end{figure}

\paragraph*{}
We demonstrate that through the dynamical valley Zeeman effect, the FMR can be optically probed by the focused laser beam illuminating the heterostructure at normal incidence. The left-handed ($\sigma_+$) and right-handed ($\sigma_-$) circularly polarized light mainly address the excitons in K and K$'$ valleys, respectively. The reflected photons with the same helicity are detected on the high-speed photodiode. 
\textcolor{ForestGreen}{Inherited by the long magnon lifetime in YIG, the quality factor ($Q\sim10^{4}$) of the ferromagnetic resonance enhances the magnon-induced Zeeman shift and the resulting signal.} Figure~\ref{figure2}\textbf{c} presents the optically-detected FMR spectra of a multilayer MoSe$_2$ flake (for a trilayer response, see SI), where the complex-valued modulation signals $R_+[\omega]$ and $R_-[\omega]$ are acquired with $\sigma_+$ and $\sigma_-$ optical polarizations, respectively.
The amplitudes $|R_\pm[\omega_m]|$ for both optical polarization, addressing K and K$'$ valleys, are the same. Nevertheless, the two signals are phase-shifted by $\pi$ ($R_+[\omega_m]= -R_-[\omega_m]$). 
In TMD monolayers, the excitonic resonance shifts due to an out-of-plane magnetization are opposite for the two valleys K and K$'$~\cite{Scharf2017,Zutic2018}. 
\color{ForestGreen}
Through this valley-resolved magnon--exciton coupling, the information of the microwave imprinted on the magnons is transferred to the reflected visible light via the excitons.
\textcolor{Sepia}{It seems that the valley-contrasting features, which are characteristic of monolayers, are preserved even in multilayer flakes. However, as we discuss later and in SI, this valley selectivity is caused by the reflection from the bottom monolayer at the van der Waals heterointerface.}
\color{black}
\paragraph*{}
\textcolor{Blue}{In order to study the dependence of the effect on the number of layers,} we perform spatially-resolved measurements over different flakes. The laser spot position on the heterostructure is controlled by a three-axis stage supporting the optical microscope. An optical micrograph of the flakes under scrutiny is presented in Fig.~\ref{figure3}\textbf{a}(i), accompanied by a topography measurement realized with an atomic force microscope shown in Fig.~\ref{figure3}\textbf{a}(ii). We define the differential optical reflectance $\Delta R[\omega] = |R_+[\omega] - R_-[\omega]|$, with $R_\pm[\omega]$ the modulation signals of the reflection originating from $\sigma_\pm$ optical polarizations, such that their $\pi$-phase difference is highlighted. 
Figure~\ref{figure3}\textbf{b} presents $\Delta R[\omega]$ along the vertical section shown in Fig.~\ref{figure3}\textbf{a}(i). This measurement shows a strong modulation of the reflected light when the laser illuminates ultra-thin MoSe$_2$ flakes. 

Dynamic effects on thinner flakes are actually belittled as the measured signal is proportional to the static optical reflection coefficient $r_\mathrm{N_L}$, with $N_L$ the number of MoSe$_2$ layers (see SI). 
To underline the dynamic response, we model the local reflectance and examine $\Delta R[\omega_m]/r_\mathrm{N_L}$ (Fig.~\ref{figure3}\textbf{d}).
The observed dynamic effects decay with the number of layers, in a fashion similar to the fraction of light coming from the very bottom layer (see SI). This qualitatively indicates that the magnon--exciton coupling originates mainly from interfacial exchange interactions~\cite{Zutic2018}. The tail at large $N_L$ may be in part attributed to the long-range effect of the tenuous dipolar field created by the magnons. \textcolor{Sepia}{As the static reflection coefficient $r_\mathrm{N_L}$ and the portion of light coming back from the bottom layer, respectively increases and decreases with an increasing number of MoSe$_2$ layers, there is an optimum number of layers giving the largest reflected optical intensity at the FMR frequency ($N_L \sim 6$ in our experimental conditions, see SI).}

\paragraph*{}
Finally, in order to get additional insights into the microscopic origin of the interaction, we quantitatively determine the magnon--exciton coupling rate $g$.
We perform calibrated measurements of the magnon-induced excitonic resonance shifts. The calibration procedure consists in comparing the optical reflection modulations induced by a known number of magnons $n_\mathrm{magnon}$ and those induced by a modulation of the laser frequency itself with a known modulation depth, without driving any magnons (see SI). The number of magnons $n_\mathrm{magnon}$ in the concerned magnetostatic mode is determined through the analysis of the FMR absorption spectra (see SI).

\begin{figure}
\includegraphics[scale=1]{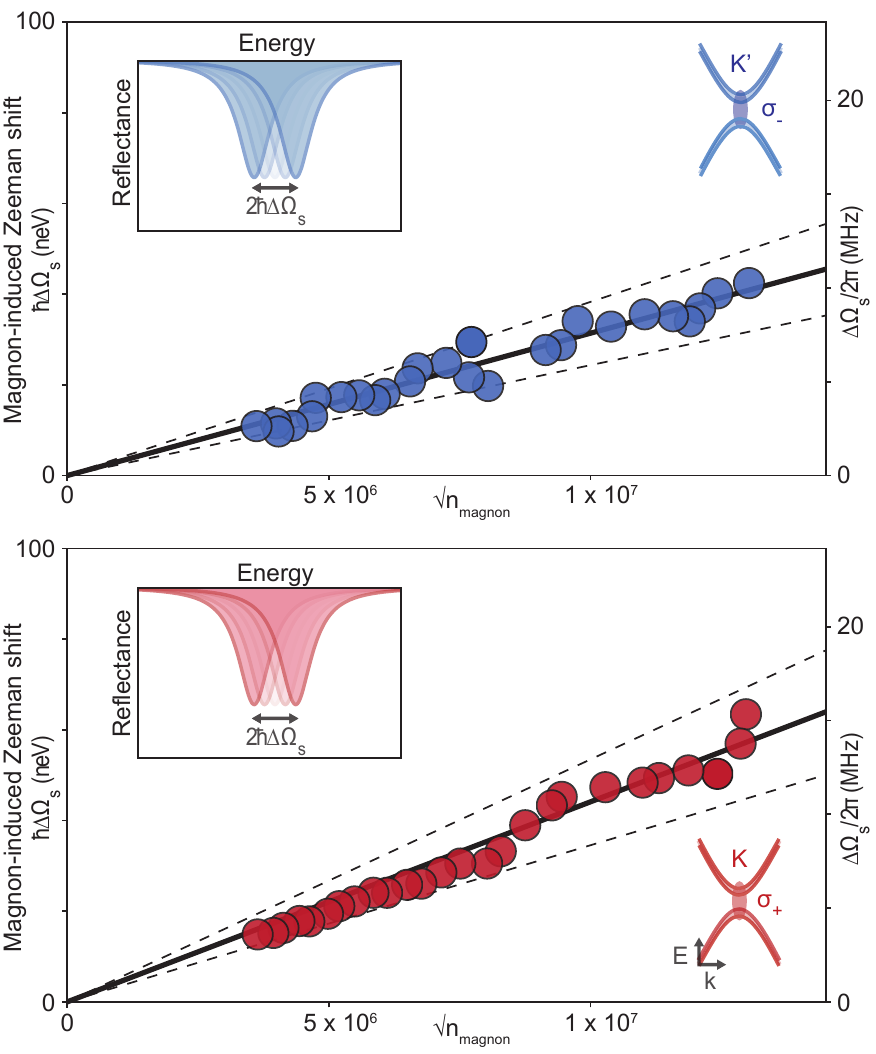}
\caption{\label{figure4} \textbf{Magnon--exciton coupling strength.} 
Evolution of the magnon-induced valley Zeeman shift $\hbar \Delta \Omega_s$ with the number of magnons $n_\mathrm{magnon}$ for a few-layer flake ($N_L \sim 8$, marked by a circle on Fig.~3\textbf{a}) for \textcolor{black}{$\sigma_{-}$ and $\sigma_{+}$} optical polarization on the upper and lower panels, respectively. 
The thick solid lines correspond to linear fits, leading to magnon--exciton coupling strengths of $\hbar g_- = (3.1 \pm 0.7) \times 10^{-15}$\,eV and $\hbar g_+=(4.4 \pm 0.9) \times 10^{-15}$\,eV for the K$'$ and K valleys, respectively, exceeding the estimated value for a coupling originating purely from dipolar effects $\hbar g_\mathrm{D} = 1.2 \times 10^{-17}\,$eV\textcolor{Sepia}{, suggesting that exchange interactions are at play} (see SI).
}
\end{figure}

The dynamical valley Zeeman shift \textcolor{Sepia}{collectively enhanced by $\sqrt{n_\mathrm{magnon}}$~\cite{Dicke54}} is $\Delta\Omega_s = g \sqrt{n_\mathrm{magnon}}$ (see SI). We measure calibrated exciton resonance shifts while ramping the microwave excitation power pumping the magnons. The coupling rate $g$ is obtained by extrapolating the excitonic resonance shifts induced by a single magnon. 
The measurement presented in Fig.~\ref{figure4} is realized on an 8-layer flake. The calibrated magnon--exciton coupling strengths are $\hbar g_+=(4.4 \pm 0.9) \times 10^{-15}$\,eV and $\hbar g_- = (3.1 \pm 0.7) \times 10^{-15}$\,eV  for left- and right-handed circular polarizations \textcolor{ForestGreen}{(equivalently $g_\pm/2\pi \sim 1\,$Hz)}. These similar $g$ values reflect that the excitons in the valleys K and K$'$ have the same sensitivity to the dynamic magnetic field, as it could be expected for a static magnetization in the plane of the film~\cite{Scharf2017}.  

The Zeeman shift induced by the effective magnetic field generated by a single magnon $B_{1\mathrm{m}}$ can be written $\hbar g= \Delta_\mu\, B_{1\mathrm{m}}$, with $\Delta_\mu$ the Zeeman shift of the exciton ($\Delta_\mu= 0.12$\,meV/T for MoSe$_2$ monolayers~\cite{Li2014}). 
We evaluate the dipolar magnetic field produced by a single magnon, small but non-zero for a finite-size sample, as $0.1\,$pT and the resultant  dipolar-originated magnon--exciton coupling strength as $\hbar g_\mathrm{D} = 1.2\times 10^{-17}\,$eV (\textcolor{ForestGreen}{$g_D/2\pi \sim 3\,$mHz}, see SI). This value constitutes an upper limit for the dipolar contribution to the effective magnon--exciton coupling strength. 
Finding magnon--exciton coupling rates $g \gg g_\mathrm{D}$ adds another evidence that the exchange interaction at these van der Waals heterointerfaces is the dominant cause of the dynamical valley Zeeman effect we observe.

\paragraph*{}
In conclusion, we have demonstrated qualitatively and quantitatively the magnon--exciton coupling at a heterointerface formed by an atomically-thin semiconductor and a magnetic film by dynamic proximity effects. 
This hybrid system allows a control of the excitonic resonances at microwave frequencies at room temperature.
Reducing the size of the magnetic film will confine the magnetic energy and enhance the magnetization oscillation amplitude per single magnon, leading to a stronger magnon--exciton coupling \color{ForestGreen} by 3--4 orders of magnitude (see SI). By optimizing the sample preparation methods and improving the interface quality, the proximity coupling may be further increased. \color{black}
Our work initiates the investigation of dynamic magnetic proximity effects at van der Waals 
heterointerfaces~\cite{Shi2019,MacNeill2019,Ciorciaro2020} towards the dynamical local control of the excitons properties, through exotic spin textures for example~\cite{Sluka2019}, valley-dependent spin transport~\cite{Ominato2020} and novel microwave-to-optics transducers~\textcolor{BurntOrange}{\cite{Hisatomi2016,Lambert2020,Lauk2020}}, establishing multiple promising routes for interconnecting efficiently optics and spin physics.

\subsection*{Acknowledgements}
\paragraph*{}
We thank F. Fu, D. Lachance-Quirion, Y. Tabuchi, S. Daimon, T. Ideue, M. Ueda, L. Kaunitz, S. Berciaud for fruitful interactions, and P.I. Sund for carefully reading the manuscript. 
This work is supported by JSPS KAKENHI (Grant No.~19K14626 and No.~19H05602) and by JST ERATO project (Grant No.~JPMJER1601). 
\paragraph*{}
The authors declare no competing interests. 
The data that support the findings of this study are available from the corresponding authors upon reasonable request.

\subsection*{Methods}
\paragraph*{Sample fabrication and characterization --}
The magnetic substrate is a commercially-available $10\,\mu$m-thick $5\times5$\,mm$^2$ YIG film grown on GGG. We performed a Piranha etching for 5\,minutes followed by an O$_2$ plasma etching for 2\,minutes on the magnetic film\color{ForestGreen}~\cite{Jungfleisch2013}\color{black}. Single crystals of MoSe$_2$ were grown via a chemical vapor transport technique\color{ForestGreen}~\cite{Shi2015}\color{black}. MoSe$_2$ flakes with typical lateral dimensions of a few micrometers were exfoliated directly onto PDMS (polydimethylsiloxane). In a glovebox filled with N$_2$, we then transferred MoSe$_2$ flakes onto the YIG film, using an all-dry transfer method\color{ForestGreen}~\cite{Castellanos2014}\color{black}, and annealed the heterostructure at $250^\circ$C for 3 hours. 

The thickness and number of layers of the flakes were determined via optical contrasts, photoluminescence spectroscopy, and atomic force microscopy images. The thickness measured by AFM is calibrated by the thickness of the monolayer flakes. The error bars in Fig.~\ref{figure3}\textbf{d} correspond to uncertainties in the determination of the thickness of the flake at the laser position considering the finite size of the laser spot, mapped on the AFM topography, extending horizontally from the lower to the higher thickness within this interval. \color{ForestGreen} These error bars translate vertically through the dependence of $r_\mathrm{N_L}$ on the number of layers $N_L$. \color{black} The fit in Fig.~\ref{figure3}\textbf{d} is performed considering the points with uncertainty on their thickness below three layers, marked as non-translucent. \color{ForestGreen}It models the portion of light coming back from the very bottom layer, evidencing the short-range magnon--exciton coupling and \textcolor{BurntOrange}{an offset} representing the \textcolor{Blue}{long-range dipolar-originated magnon--exciton coupling} and the \textcolor{BurntOrange}{instrumental} detection background (see SI). The point reported at $t\sim20\,$nm in Fig.~3\textbf{d} encompasses different terraces and cannot be normalized properly. 

The calibration in Fig.~4 is performed on a flake which has a constant thickness on an area much larger than the laser spot size. The difference in the coupling strength between left and right-handed circular polarizations seems not be substantial and most likely due to systematic errors such as the quality of the polarizing cubes and electrical environment. \color{black}

\paragraph*{Optical setup --}
Before being focused by a microscope objective, the linearly-polarized laser beam passes through a quarter-wave plate to be turned into left- ($\sigma_+$) or right-handed ($\sigma_-$) circularly polarized light and mainly address the excitons in K and K$'$ valleys, respectively. The light reflected off the sample goes through the quarter-wave plate on the way back and is filtered out by a polarizing beamsplitter, such that only the reflected photons sharing the same helicity with the incident photons are directed towards a fiber-coupled high-speed photodiode (bandwidth: 12\,GHz). The electric signal of the photodiode is analyzed on a network analyzer (Figs.~1--3) and a spectrum analyzer (Fig.~4, see SI). 

\bibliography{biblio_formate}

\end{document}


\title{Supplementary Information: Magnon-exciton proximity coupling at a van der Waals heterointerface}
\author{A. Gloppe}
\email[]{arnaud.gloppe@ipcms.unistra.fr}
\affiliation{Universit\'e de Strasbourg, CNRS, Institut de Physique et Chimie des Mat\'eriaux de Strasbourg (IPCMS), UMR 7504, F-67000 Strasbourg, France}
\affiliation{Research Center for Advanced Science and Technology (RCAST), The University of Tokyo, Meguro-ku, Tokyo 153-8904, Japan}
\author{M. Onga}
\affiliation{Quantum-Phase Electronics Center (QPEC) and Department of Applied Physics, The University of Tokyo, Tokyo 113-8656, Japan}
\author{R. Hisatomi}
\affiliation{Research Center for Advanced Science and Technology (RCAST), The University of Tokyo, Meguro-ku, Tokyo 153-8904, Japan}

\author{A. Imamo\u{g}lu}
\affiliation{Institute for Quantum Electronics, ETH Zurich, CH-8093, Zurich, Switzerland}

\author{Y. Nakamura}
\affiliation{Research Center for Advanced Science and Technology (RCAST), The University of Tokyo, Meguro-ku, Tokyo 153-8904, Japan}
\affiliation{RIKEN Center for Quantum Computing (RQC), Wako, Saitama 351-0198, Japan}

\author{Y. Iwasa}
\email{iwasa@ap.t.u-tokyo.ac.jp}
\affiliation{Quantum-Phase Electronics Center (QPEC) and Department of Applied Physics, The University of Tokyo, Tokyo 113-8656, Japan}
\affiliation{RIKEN Center for Emergent Matter Science (CEMS), Wako, Saitama 351-0198, Japan}

\author{K. Usami}
\email{usami@qc.rcast.u-tokyo.ac.jp}
\affiliation{Research Center for Advanced Science and Technology (RCAST), The University of Tokyo, Meguro-ku, Tokyo 153-8904, Japan}

\date{\today}

\maketitle

\tableofcontents
\clearpage

\section{Theoretical model}
In this part, we provide the theoretical framework describing the interactions between optical photons and excitons, magnons and microwave photons in our hybrid system. 
\subsection{Excitons}
\subsubsection*{Interaction between light and excitons in a MoSe$_2$ monolayer}
The K excitons of a MoSe$_{2}$ flake can be modeled as excitations in bosonic modes~\cite{Zeytinoglu2017}, for which the Hamiltonian reads
\begin{equation}
H_\mathrm{ex}  = \hbar \Omega_\mathrm{ex} \hat{x}_{\mathrm{K}}^{\dagger}\hat{x}_{\mathrm{K}}, \label{eq:Ha}
\end{equation}
where $\hat{x}_{\mathrm{K}}$ and $\hat{x}_{\mathrm{K}}^{\dagger}$ are the annihilation and creation operators for K excitons and $\Omega_\mathrm{ex}/2\pi$ their resonance frequency. Similarly, the K$'$ exciton can be modeled with the Hamiltonian
\begin{equation}
H_{\mathrm{ex}'} = \hbar \Omega_\mathrm{ex} \hat{x}_{\mathrm{K}'}^{\dagger}\hat{x}_{\mathrm{K}'}.
\end{equation}

We suppose that the K excitons are selectively driven by circular-polarized light. 
\textcolor{ForestGreen}{
We focus for the moment on the situation in which there is no magnon excited in the film, in order to express the static optical reflection coefficient of a monolayer flake~\cite{Zeytinoglu2017,Scuri2018,Back2018}.} The interaction between the optical mode propagating from the laser source to the sample and the excitons can be described by~\cite{Zeytinoglu2017, Clerk2010}
\begin{align}
\label{eq:Hl}
H_{l} =& -i \hbar \sqrt{\kappa_l} \int \frac{d \Omega}{2 \pi} \left( \hat{x}_{\mathrm{K}}^{\dagger} \hat{l}[\Omega]- \hat{x}_{\mathrm{K}} \hat{l}^{\dagger}[\Omega]\right)  \\
=& -i \hbar \sqrt{\kappa_l} \left(\hat{x}_{\mathrm{K}}^{\dagger}\,l_\mathrm{in} e^{-i \Omega_{L} t} - \hat{x}_{\mathrm{K}} \, l_\mathrm{in} e^{i \Omega_{L}t}  \right)\nonumber,
\end{align}
where $\kappa_l$ represents the radiative decay rate of the excitons in the optical mode described by $\hat{l}$, $l_\mathrm{in} = \sqrt{P_\mathrm{opt}/\hbar \Omega_{L}}$ is the photon flux with $P_\mathrm{opt}$ the power and $\Omega_{L}$ the angular frequency of the drive light. Here, the quantum operators $\hat{l}^{\dagger}[\Omega]$ and $\hat{l}[\Omega]$ are replaced by their classical values with definite amplitude and phase, i.e., $l_\mathrm{in} e^{i \Omega t} \delta[\Omega-\Omega_{L}]$ and $l_\mathrm{in} e^{-i \Omega t} \delta[\Omega-\Omega_{L}]$, respectively.

\color{ForestGreen}
By analogy to the operator $\hat{l}$ associated with the driven optical mode, the operator $\hat{r}$ describes the optical field mode propagating in the opposite direction and interacts with the excitons through the Hamiltonian
\begin{equation}
\label{eq:Hr}
H_{r} = -i \hbar \sqrt{\kappa_r} \int \frac{d \Omega}{2 \pi} \left( \hat{x}_{\mathrm{K}}^{\dagger} \hat{r}[\Omega]- \hat{x}_{\mathrm{K}} \hat{r}^{\dagger}[\Omega]\right)  
\end{equation}
with $\kappa_r$ being the radiative decay rate of the excitons in this optical mode. The free Hamiltonian for both optical modes reads
\begin{equation}
H_{0} = \int \frac{d \Omega}{2 \pi} \, \hbar \Omega \left(\hat{r}^{\dagger}[\Omega] \hat{r}[\Omega] + \hat{l}^{\dagger}[\Omega] \hat{l}[\Omega]\right).
\end{equation}
\paragraph*{}
Following similar developments as in~\cite{Zeytinoglu2017} inspired by the input-output formalism in quantum optics~\cite{Walls2007, Clerk2010}, the total Hamiltonian to be considered in the rotating frame with the drive light frequency $\Omega_L/2\pi$ is
\begin{equation}
\label{eq:checkH}
\textcolor{Blue}{H_\mathrm{tot}} = H_\mathrm{ex} + H_0  + H_l + H_r.
\end{equation}
We define the input and output time-domain operators ($\hat{s} = \hat{l}$ or $\hat{r}$)
\begin{equation}
\hat{s}_\mathrm{in/out}(t) 		= \int \frac{d\Omega}{2 \pi} e^{-i \Omega \left( t-\tau	  \right)}   \hat{s}[\Omega],
\end{equation}
with $\tau < t$ for the input operator $\hat{s}_\mathrm{in}$ and $\tau > t$ for the output operator $\hat{s}_\mathrm{out}$.

The Heisenberg equation for $\hat{x}_K$ can then be written as
\begin{equation}
\dot{\hat{x}}_K = -i \Omega_\mathrm{ex} \hat{x}_K - \frac{\kappa}{2} \hat{x}_K - \sqrt{\kappa_l} \hat{l}_\mathrm{in} - \sqrt{\kappa_r} \hat{r}_\mathrm{in} 
\end{equation}
with $\kappa$ the total exciton dissipation rate which encompasses radiative and non-radiative decay channels. Its Fourier transform is
\begin{equation}
\label{eq:xomega}
\hat{x}_K[\omega] = -\frac{\sqrt{\kappa_l} \hat{l}_\mathrm{in}[\omega] + \sqrt{\kappa_r} \hat{r}_\mathrm{in}[\omega]}{i(\Omega_\mathrm{ex}-\omega)+\frac{\kappa}{2}}.
\end{equation}
The input-output relation in this system reads
\begin{align}
\hat{r}_\mathrm{out} &= \hat{r}_\mathrm{in} + \sqrt{\kappa_\mathrm{r}} \hat{x}_{\mathrm{K}}. 
\end{align}
Neglecting the contribution of the quantum light $\hat{r}_\mathrm{in}$, 
it comes
\begin{equation}
r_\mathrm{out} = -\frac{\sqrt{\kappa_\mathrm{r} \kappa_\mathrm{l}} }{i (\Omega_\mathrm{ex}-\Omega_L) +\frac{\kappa}{2}} \, l_\mathrm{in}.
\end{equation}
The term
 \textcolor{Blue}{
\begin{equation}
\label{eq:rlevrai}
r_\mathrm{1L} \equiv \frac{\sqrt{\kappa_\mathrm{r} \kappa_\mathrm{l}} }{i (\Omega_\mathrm{ex}-\Omega_L) +\frac{\kappa}{2} }
\end{equation} 
}can be viewed as the static reflection coefficient of the MoSe$_2$ monolayer at the optical frequency $\Omega_L/2\pi$. The static reflectance for a monolayer is then
\begin{equation}
|r_\mathrm{1L}|^2 = \frac{\kappa_\mathrm{r} \kappa_\mathrm{l}}{(\Omega_\mathrm{ex}-\Omega_L)^2 +\frac{\kappa^2}{4} }.
\end{equation}
If there were no non-radiative decay such that $\kappa = \kappa_l+\kappa_r$, and assuming that $\kappa_l=\kappa_r$, then $|r_\mathrm{1L}|^2 = 1$ when the excitons are optically excited close to their resonance ($\Omega_L=\Omega_\mathrm{ex}$).

\color{black}

\subsubsection*{Simple model for the optical reflection of multilayer flakes}
\label{subsec:multilayer}
\begin{figure}
\includegraphics[scale=1.1]{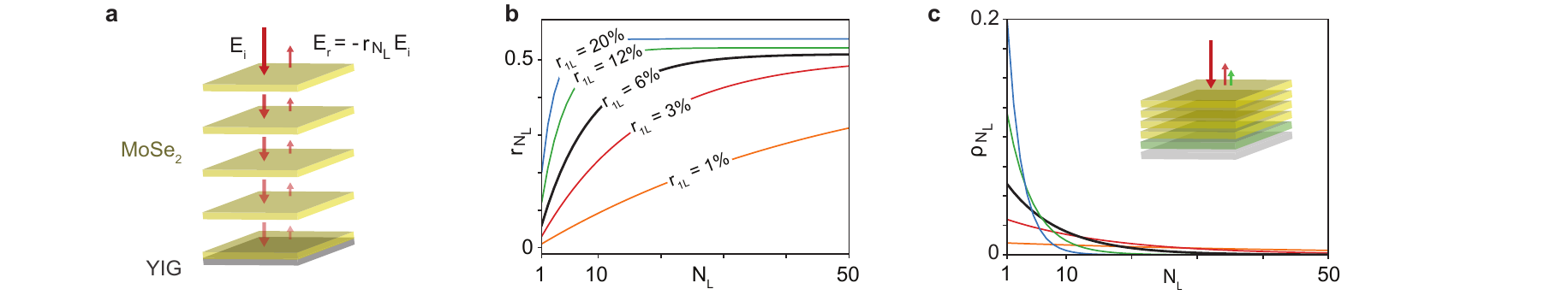}
\caption{\label{SIf:model_refl} \textbf{Model for the optical reflection of multilayer flakes.} \color{ForestGreen}\textbf{a,} Simple schematic of the model with an incoming field $E_i$ and a reflected electric field $E_r$, for a number of layer $N_L$ and an effective monolayer reflection coefficient $r_\mathrm{1L}$, illustrating Eq.~(\ref{eq:r}). \textbf{b,} Modeled reflection coefficient $r_\mathrm{N_L}$ for various $r_\mathrm{1L}$ (20\%, 12\%, 6\%, 3\% and 1\% in blue, green, black, red and orange, respectively). \textbf{c,} Relative portion $\rho_\mathrm{N_L}$ of the total reflected light coming back from the very bottom layer, highlighted in green, as predicted by Eq.~(\ref{SIeq:rhob}).\color{black}}
\end{figure}

\textcolor{ForestGreen}{In the previous paragraphs, we have expressed the static reflection coefficient \textcolor{Blue}{$r_\mathrm{1L}$} for a monolayer flake. Here we present a simple model to deduce the static reflection coefficient \textcolor{Blue}{$r_\mathrm{N_L}$} for a multilayer flake.}
Assuming that the reflectivity spectrum is not dramatically affected  by the number of layers $N_L$ when $N_L > 5$~\cite{Niu2018}, and that multiple reflections between layers \textcolor{Blue}{(as $|r_\mathrm{1L}|^2 \ll 1$)} and absorption can be neglected, the \textcolor{ForestGreen}{static} reflection coefficient $r_\mathrm{N_L}$ for a multilayer flake can be modeled as
\begin{equation}
\label{eq:r}
r_\mathrm{N_L} = r_\mathrm{1L} \sum_{k=0}^{N_L-1} (1-r_\mathrm{1L})^{2k}.
\end{equation}
\color{ForestGreen} The reflection from the magnetic film has been omitted for simplicity. The model is illustrated in Fig.~\ref{SIf:model_refl}\textbf{a}. The static reflection coefficient $r_\mathrm{N_L}$ for a given value of $r_\mathrm{1L}$ is plotted in Fig.~\ref{SIf:model_refl}\textbf{b}. 
We can then estimate the relative portion of light coming from the very bottom layer
\color{ForestGreen}
\begin{equation}
\label{SIeq:rhob}
\rho_{N_L} = r_\mathrm{1L}(1-r_\mathrm{1L})^{2(N_L-1)},
\end{equation}
decaying rapidly with the number of layers as plotted in Fig.~\ref{SIf:model_refl}\textbf{c}. 
\textcolor{Blue}{The experimental determination of the effective $r_\mathrm{1L}$ and the dependence of the apparent dynamical perturbation on the number of layers are shown in Sec.~\ref{subsec:drfNL}.}
\color{black}
\subsection{Magnons}
\subsubsection*{Landau-Lifshitz equations}
\textcolor{ForestGreen}{We describe here the dynamics of the magnons in the magnetostatic mode.} The magnetic film plane is normal to the $z$-axis and the static magnetic field $H_\mathrm{DC}$ is along the $y$-axis. 
The lateral dimension of the film $w$ = 5\,mm being much larger than its thickness $d$ = 10\,$\mu$m, the transverse magnetization components $(m_x,m_z)$ follow the Landau-Lifshitz equations~\cite{Stancil2009}:
\begin{eqnarray}
\dot{m}_{x} &=& \gamma \mu_{0} \left( H_\mathrm{DC} + M_{s} \right) m_{z} \\
\dot{m}_{z} &=& -\gamma \mu_{0} H_\mathrm{DC} \,  m_{x}
\end{eqnarray}
\textcolor{Blue}{where $\mu_0$ is the magnetic constant, $\gamma$ is the gyromagnetic ratio and $M_s$ is the saturation magnetization of the film}. 

\subsubsection*{Heisenberg-Langevin equation}
\textcolor{ForestGreen}{We define the ladder operators describing the magnons in the magnetostatic mode}. By working with the scaled transverse magnetization components 
\begin{eqnarray}
\tilde{m}_{x} &=& \frac{m_{x}}{\sqrt{\frac{H_\mathrm{DC}+M_{s}}{2H_\mathrm{DC}+M_{s}}}} \label{eq:mx}\\
\tilde{m}_{z} &=& \frac{m_{z}}{\sqrt{\frac{H_\mathrm{DC}}{2H_\mathrm{DC}+M_{s}}}}, \label{eq:mz}
\end{eqnarray}
we have 
\begin{eqnarray}
\label{eq:mxmz_1}
\dot{\tilde{m}}_{x} &=& \omega_{m} \tilde{m}_{z}, \\
\label{eq:mxmz_2}
\dot{\tilde{m}}_{z} &=& -\omega_{m} \tilde{m}_{x},
\end{eqnarray}
\textcolor{Blue}{with $\omega_m/2\pi$ the FMR frequency such that
\begin{equation}
\label{eq:freqmagnon}
\omega_m = \mu_0 \gamma \sqrt{H_\mathrm{DC}\left(H_\mathrm{DC}+M_s\right)}.
\end{equation} 
}The normal modes can then be described by 
\begin{align}
\tilde{m}_{-} &= \tilde{m}_{z} -i \tilde{m}_{x} \label{eq:m-}\\
\tilde{m}_{+} &= \tilde{m}_{z} +i \tilde{m}_{x}. \label{eq:m+}
\end{align}
Here, $\tilde{m}_{-}$ and $\tilde{m}_{+}$ represent the rescaled circular movements of the transverse magnetization oscillations. Those are represented in terms of the creation and annihilation operators~\cite{Holstein1940} as
\begin{align}
\tilde{m}_{-}(t) &= \sqrt{2}\, \frac{M_s}{\sqrt{N}}  \hat{a}(t), \label{eq:a} \\
\tilde{m}_{+}(t) &= \sqrt{2}\, \frac{M_s}{\sqrt{N}}  \hat{a}^\dagger(t), \label{eq:adag}
\end{align}
where $N$ is the total number of spins in the film. 
\color{ForestGreen}
The equation of motion for the magnons in the magnetostatic mode given in Eqs.~(\ref{eq:mxmz_1}-\ref{eq:mxmz_2}) 
can then be rewritten as
\begin{equation} 
\dot{\hat{a}}(t) = -i \omega_{m} \hat{a}(t). \label{eq:eomQ0}
\end{equation}
\color{black}
The Hamiltonian for the magnetostatic mode then reads
\begin{equation}
H_{m} = \hbar \omega_{m} \hat{a}^{\dagger} \hat{a}. \label{eq:Hm}
\end{equation}
\color{ForestGreen}
\paragraph*{} 
By phenomenologically adding damping and external coupling terms, \color{black} the equation of motion for the magnons in the magnetostatic mode can be written as a standard Heisenberg-Langevin equation:
\begin{equation} 
\dot{\hat{a}}(t) = -i \omega_{m} \hat{a}(t) - \frac{\gamma_m}{2} \hat{a}(t) - \sqrt{\gamma_{e}} \hat{a}_{\mathrm{in}}, \label{eq:eomQ}
\end{equation}
where $\hat{a}_{\mathrm{in}}$ is the annihilation operator for the microwave field in the transmission line connected to the microwave loop-antenna. The magnon decay rate is $\gamma_m = \gamma_{0} + \gamma_{e}$, where $\gamma_{0}$ the intrinsic damping rate and $\gamma_{e}$ the external coupling rate between the magnetostatic mode and the antenna. \textcolor{Blue}{Derived from Eq.~(\ref{eq:eomQ}), the $S_{11}$ parameter for the transmission line then reads:
\begin{equation}
\label{eq:S11}
S_{11}[\omega] = \dfrac{i(\omega-\omega_m) - \frac{1}{2}(\gamma_0-\gamma_e)}{i(\omega-\omega_m) - \frac{1}{2} (\gamma_0+\gamma_e)}.
\end{equation}
}
\subsection{Magnon-exciton coupling}
\label{sec:theory}
\textcolor{ForestGreen}{
In this section, we develop the formalism describing the coupling between magnons and excitons, defining the magnon-exciton coupling rate $g$ and expressing the magnon-induced Zeeman shift $\Delta \Omega_s$.}

\subsubsection*{Interaction between magnons and excitons}
The excitons in a MoSe$_2$ monolayer and the magnons excited in the YIG film interact at the heterointerface through a valley Zeeman effect. 
\textcolor{ForestGreen}{
The Zeeman energy shift $\hbar\tilde{g}$ induced by the magnons on the excitons can be written in generic form as 
\begin{equation}
\hbar\tilde{g} = \mathrm{g}_\mathrm{ex} \mu_\mathrm{B} \,B_m, \label{eq:gtilde}
\end{equation}
where $g_{\mathrm{ex}}$ is the effective g-factor of the excitons in MoSe$_2$~\cite{Koperski2018}, $\mu_{\mathrm{B}}$ is the Bohr magneton, and $B_m$ is the magnetic field originating from the magnons and felt by the excitons. This magnetic field $B_m$ arises from the transverse magnetization $m_z$ due to magnons which, by combining Eqs.{~(\ref{eq:mz}-\ref{eq:adag})}, reads}
\color{black}
\begin{equation}
m_{z} = \sqrt{\frac{H_\mathrm{DC}}{2 H_\mathrm{DC} + M_s}} \sqrt{2}\, \frac{M_{s}}{\sqrt{N}} \left( \frac{\hat{a}+\hat{a}^{\dagger}}{2} \right). \label{eq:mz2}
\end{equation}
\textcolor{ForestGreen}{
Note that the magnon-induced Zeeman shift $\hbar\tilde{g}$ and the equivalent Zeeman field $B_m$ are proportional to $1/\sqrt{N}$, reflecting the fact that the magnetostatic mode is delocalized over the entire sample volume $V$. The magnetic field $B_m$ felt by the excitons could stem from a short-range interfacial exchange field or from a long-range dipolar field.}
The magnon--exciton interaction Hamiltonian is then given by 
\begin{equation}
H_{i} = \textcolor{ForestGreen}{+\hbar \tilde{g} \,  \hat{x}_{\mathrm{K}}^{\dagger}\hat{x}_{\mathrm{K}} = } +\hbar g \left( \frac{\hat{a} + \hat{a}^{\dagger}}{2} \right) \hat{x}_{\mathrm{K}}^{\dagger}\hat{x}_{\mathrm{K}}, \label{eq:Hai}
\end{equation}
for the K exciton, and 
\begin{equation}
H_{i'} = \textcolor{ForestGreen}{-\hbar \tilde{g} \,  \hat{x}_{\mathrm{K}}^{\dagger}\hat{x}_{\mathrm{K}} = }  -\hbar g \left( \frac{\hat{a} + \hat{a}^{\dagger}}{2} \right) \hat{x}_{\mathrm{K}'}^{\dagger}\hat{x}_{\mathrm{K}'}, \label{eq:Hai}
\end{equation}
for the K$'$ exciton, where the opposite sign reflects that the K$'$ exciton is the time-reversal partner of the K exciton.
\textcolor{ForestGreen}{
The magnon--exciton coupling rate $g$ corresponds to the Zeeman shift induced by a single magnon. 
From Eqs.~(\ref{eq:gtilde}-\ref{eq:mz2}), we note that $g$ is also proportional to $1/\sqrt{N}$. We note that the form of $g$ is generic and valid whatever the origin of the coupling might be.}

In a bare multilayer flake, the inversion symmetry may not be broken, such that the valley degree of freedom is not well-defined. 
However, \textcolor{Blue}{if the magnetic field $B_{m}$ is due to interfacial interactions, the valley Zeeman shift and the resultant magnon-exciton coupling predominantly coming from the bottom layer of the multilayer flake, the interfacial exchange field might be responsible for breaking the inversion symmetry and preserving the associated optical selection rules.}

\paragraph*{}
\textcolor{ForestGreen}{Omitting in Eq.~(\ref{eq:checkH}) the terms $H_r$ and $H_0$ for the sake of clarity, the total Hamiltonian in the laboratory frame reads}
\begin{equation}
H = H_{m} + H_\mathrm{ex} + H_{i} + H_{l}. \label{eq:H}
\end{equation}
Let us introduce a displacement operator for the K exciton
\begin{equation*}
D(\zeta) = e^{\zeta \hat{x}_{\mathrm{K}}^{\dagger} - \zeta^{\ast} \hat{x}_{\mathrm{K}}}, 
\end{equation*}
with the displacement amplitude $\zeta$ to eliminate the drive term as \textcolor{ForestGreen}{
\begin{equation}
\dot{\zeta} = -i \Omega_\mathrm{ex} \zeta - \frac{\kappa}{2} \zeta +  \sqrt{\kappa_l} \ l_\mathrm{in} e^{-i\Omega_{L}t}. \label{eq:EOMa}
\end{equation}}

\paragraph*{}
The displacement amplitude reads
\textcolor{ForestGreen}{ 
\begin{equation}
\zeta = \sqrt{n_\mathrm{ex}} e^{i\varphi_\mathrm{ex}}\, e^{-i\Omega_{L}t} \label{eq:Dex}
\end{equation}
with
\begin{equation}
\label{eq:nex}
n_\mathrm{ex} = \frac{\kappa_l}{\left( \Omega_\mathrm{ex} - \Omega_L  \right)^2 + \frac{\kappa^2}{4}} \frac{P_\mathrm{opt}}{\hbar \Omega_{L}}, 
\end{equation}
which can be viewed as the effective exciton number optically driven and $\varphi_\mathrm{ex}$ a phase term depending on $(\Omega_\mathrm{ex}-\Omega_L)/\kappa$.}
\paragraph*{}
The total Hamiltonian $H$ defined in Eq.~(\ref{eq:H}) can be unitary-transformed with $D(\zeta)$ to obtain
\begin{eqnarray}
\tilde{H} &=& D^{\dagger}(\zeta) H D(\zeta) + i\hbar \dot{D}^{\dagger}(\zeta) D(\zeta) \nonumber\\
&=& \hbar \omega_{m} \hat{a}^{\dagger} \hat{a} + \left( \hbar \Omega_\mathrm{ex}  + \hbar g \left( \frac{\hat{a} + \hat{a}^{\dagger}}{2} \right) \right) \hat{x}_{\mathrm{K}}^{\dagger} \hat{x}_{\mathrm{K}} + \hbar g \left( \frac{\hat{a} + \hat{a}^{\dagger}}{2} \right)  \left( \zeta \hat{x}_{\mathrm{K}}^{\dagger} + \zeta^{\ast} \hat{x}_{\mathrm{K}} \right) \label{eq:H2D}
\end{eqnarray}
where all the c-numbered terms are omitted as being mere energy offsets.

Finally, by assuming to be optically close to resonance and performing another unitary transformation  
\begin{equation}
U = e^{i \frac{t}{\hbar} \left( \hbar \Omega_{L} \hat{x}_{\mathrm{K}}^{\dagger} \hat{x}_{\mathrm{K}} \right) }, \label{eq:Ux}
\end{equation} 
which transforms the exciton operators $\hat{x}_{\mathrm{K}}$ and $\hat{x}_{\mathrm{K}}^{\dagger}$ to $\hat{x}_{\mathrm{K}}e^{-i \Omega_{L}t}$ and $\hat{x}_{\mathrm{K}}^{\dagger}e^{i\Omega_{L}t}$, respectively,  we have 
\begin{equation}
\mathcal{H} = \hbar \omega_{m} \hat{a}^{\dagger} \hat{a} + \left(\hbar \left( \Omega_\mathrm{ex} -\Omega_{L} \right) + \underbrace{\hbar g \left( \frac{\hat{a} + \hat{a}^{\dagger}}{2} \right)}_{\mathrm{energy \ shift}} \right)\hat{x}_{\mathrm{K}}^{\dagger} \hat{x}_{\mathrm{K}} + \underbrace{\hbar g \sqrt{n_\mathrm{ex}} \left( \frac{\hat{a} + \hat{a}^{\dagger}}{2} \right)}_{\mathrm{displacement}} \left( \hat{x}_{\mathrm{K}}^{\dagger}+ \hat{x}_{\mathrm{K}} \right). \label{eq:H4D}
\end{equation}
We can see that the magnon--exciton coupling characterized by the rate $g$ manifests itself within two terms. From the view point of the exciton, these are the \textit{energy shift} term (quadratic term) and the \textit{displacement} term (linear term)\textcolor{ForestGreen}{, both of which are dependent on the $z$-component of the transverse magnetization $m_z$ given in Eq.~(\ref{eq:mz2})}.

\subsubsection*{Magnon-induced modulation of the exciton resonance frequency}
\label{ss:magnonZeeman}
We neglect the magnon-induced displacement term since it is typically far smaller than the optically-induced displacement term. 
Referring to Eq.~(\ref{eq:eomQ}), driving the magnons resonantly by an external microwave field at a frequency $\omega_{d}/2\pi$ through a microwave loop-antenna leads to the displacement of $\hat{a}$ to 
\textcolor{ForestGreen}{$\hat{a}+\sqrt{n_\mathrm{magnon}}e^{i\varphi_m}\, e^{-i \omega_{d}t}$, where
\begin{equation}
\label{eq:nmagnongen}
n_\mathrm{magnon} = \dfrac{\gamma_e}{(\omega_m-\omega_d)^2+\frac{\gamma_m^2}{4}} \dfrac{P_\mathrm{MW}}{\hbar\omega_d},
\end{equation}
}
with $P_\mathrm{MW}$ the microwave drive power such that $\sqrt{P_\mathrm{MW}/\hbar \omega_{d}} = \langle \hat{a}_\mathrm{in} \rangle$, and $\varphi_m$ a phase depending on $(\omega_m-\omega_d)/\gamma_m$.

Neglecting the quantum fluctuations and retaining only the classically-driven part close to resonance ($\omega_d \sim \omega_m$), the magnon-induced energy shift term in Eq.~(\ref{eq:H4D}), $\hbar g \left( \frac{\hat{a} + \hat{a}^{\dagger}}{2} \right) \hat{x}_{\mathrm{K}}^{\dagger}\hat{x}_{\mathrm{K}}$, thus becomes $\hbar g \left(\sqrt{n_\mathrm{magnon}} \cos \omega_m t\right) \, \hat{x}_{\mathrm{K}}^{\dagger} \hat{x}_{\mathrm{K}}$, modulating the exciton resonance angular frequency $\Omega_\mathrm{ex}$ as 
\begin{equation}
\label{eq:omexmod}
\Omega_\mathrm{ex}(t) = \Omega_\mathrm{ex} + \Delta \Omega_s \cos \omega_m t
\end{equation}
with 
\begin{equation}
\Delta \Omega_s = g \sqrt{n_\mathrm{magnon}}
\end{equation} the magnon-induced Zeeman shift. 
\textcolor{ForestGreen}
{
\paragraph*{}
From Eq.~(\ref{eq:rlevrai}), for a small Zeeman shift ($\Delta \Omega_s \ll \Omega_\mathrm{ex}$), this modulation of the excitonic resonance turns into a modulation of the reflection coefficient of the flake as 
\begin{align}
r_\mathrm{1L}(t) =  \frac{\sqrt{\kappa_r \kappa_l}}{i(\Omega_\mathrm{ex}(t) -\Omega_L)+\frac{\kappa}{2}} = r_\mathrm{1L} + \delta r_\mathrm{1L}(t)
\end{align}
with the reflected field $E_r = - r_\mathrm{1L}(t) E_i$,
where 
\begin{equation}
\label{deltarwm}
\delta r_\mathrm{1L}(t) \equiv \underbrace{\frac{\textcolor{Blue}{-}i \sqrt{\kappa_l\kappa_r}}{\left[i(\Omega_\mathrm{ex}-\Omega_L)+\frac{\kappa}{2}\right]^2}}_{\frac{\partial r_\mathrm{1L}}{\partial(\Omega_\mathrm{ex}-\Omega_L)}} \Delta\Omega_s \cos \omega_m t.
\end{equation}
}
\subsubsection*{Generalization to multilayer flakes}
\label{sec:genmultiflake}
\textcolor{Blue}{
For a flake of $N_L$ layers, the reflection coefficient can be generalized as $r_\mathrm{N_L}(t) = r_\mathrm{N_L} + \delta r_\mathrm{N_L}(t)$,} \color{Sepia}such that
\begin{equation}
E_r = - r_\mathrm{N_L}(t) E_i
\end{equation}
with $\delta r_\mathrm{N_L}$ the apparent dynamical perturbation of the reflection coefficient written as 
\begin{equation}
\label{deltarNLwm}
\delta r_\mathrm{N_L}(t) = \delta r_\mathrm{N_L} \, \cos \omega_m t  
\end{equation} 
where $\delta r_\mathrm{N_L}$ is a time-constant encapsulating the microscopic mechanisms of the magnon--exciton coupling in multilayer flakes. The optical intensity of the reflected beam is then naturally proportional to 
\begin{align*}
|E_r|^2 &= |r_\mathrm{N_L} + \delta r_\mathrm{N_L}(t)|^2 |E_i|^2\\
&= |r_\mathrm{N_L}^2 + 2 r_\mathrm{N_L} \delta r_\mathrm{N_L} \cos \omega_m t+ \delta r_\mathrm{N_L}^2 \cos^2 \omega_m t | |E_i|^2,
\end{align*}
such that in particular, its Fourier component at the FMR frequency $\omega_m/2\pi$ reads
\begin{equation}
\label{eq:Erwom}
|E_r|^2[\omega_m] \propto |r_\mathrm{N_L} \,\, \delta r_\mathrm{N_L}|.
\end{equation}

Assuming that the magnon--exciton coupling is dominated by exchange interactions between the very bottom layer of MoSe$_2$ and the YIG substrate, then the modification of the reflection coefficient follows, as a function of the number of layers, the portion of light $\rho_\mathrm{N_L}$ coming back from the very bottom layer as modeled in Eq.~(\ref{SIeq:rhob}), i.e. $\delta r_\mathrm{N_L} \propto \rho_\mathrm{N_L}$ and $|E_r|^2[\omega_m] \propto r_\mathrm{N_L}  \rho_\mathrm{N_L}$.
\begin{figure}[h]
\includegraphics[scale=1]{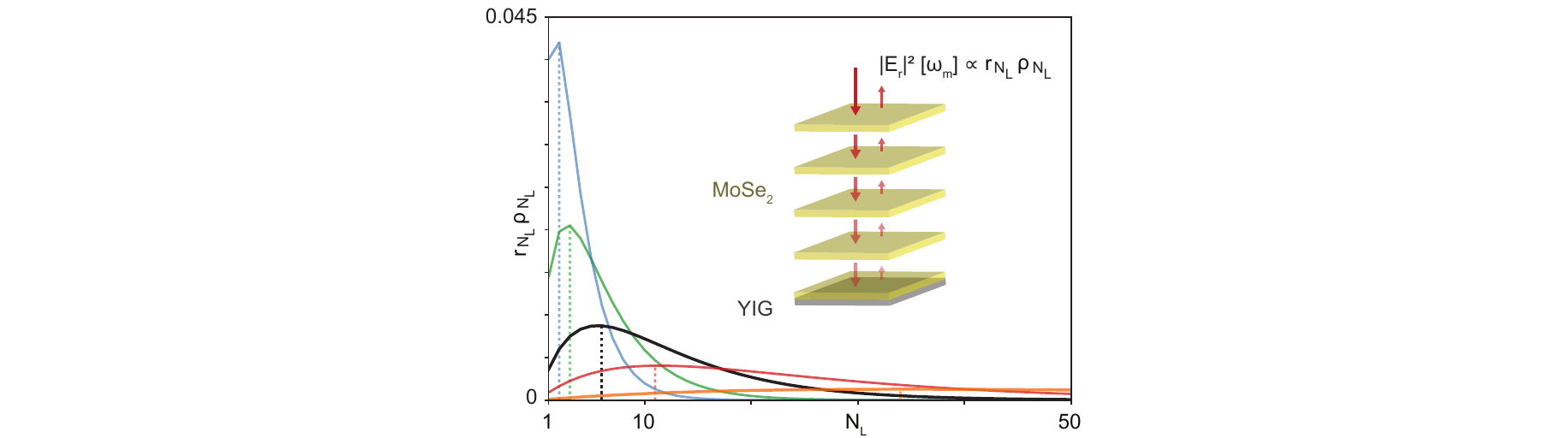}
\caption{\label{SIf:reflpartdynA} \textcolor{Sepia} {Modeled evolution $r_\mathrm{N_L} \rho_\mathrm{N_L}$, proportional to the Fourier component of the reflected optical intensity $|E_r|^2[\omega_m]$, with the number of layers and the effective monolayer reflection coefficient ($r_\mathrm{1L}$: 20\%, 12\%, 6\%, 3\% and 1\% in blue, green, black, red and orange, respectively), following Eqs.~(\ref{eq:r}-\ref{SIeq:rhob}). The number of layers giving the largest optical signal at the FMR frequency is marked with a dotted line for each $r_\mathrm{1L}$.}} 
\end{figure} 
With an increasing number of layers, $r_\mathrm{N_L}$ and $\rho_\mathrm{N_L}$ increases and decreases, respectively, leading to an optimum number of layers at a given $r_\mathrm{1L}$ inducing the largest reflected optical signal at the FMR frequency. We illustrate this trade-off by plotting the theoretical values of $r_\mathrm{N_L} \rho_\mathrm{N_L}$ along our model in Fig.~\ref{SIf:reflpartdynA}, following Eqs.~(\ref{eq:r}-\ref{SIeq:rhob}), in agreement with the experimental data presented later in Fig.~\ref{SIf:reflpartdynB}.
\color{black}

\section{Experimental determination of the magnon--exciton coupling strength}
\label{sec:g}
The phenomenological magnon--exciton coupling rate $g$ captures the essence of the interaction. Evaluating $g$ provides a useful insight into the microscopic origin of the magnon--exciton coupling. Here, we show in detail the calibration procedure to determine the magnon-induced Zeeman shifts $\Delta\Omega_s$ and the magnon--exciton coupling rate $g$ presented in the main text. 
\subsection{Setup}
\begin{figure}[h]
\includegraphics[scale=1]{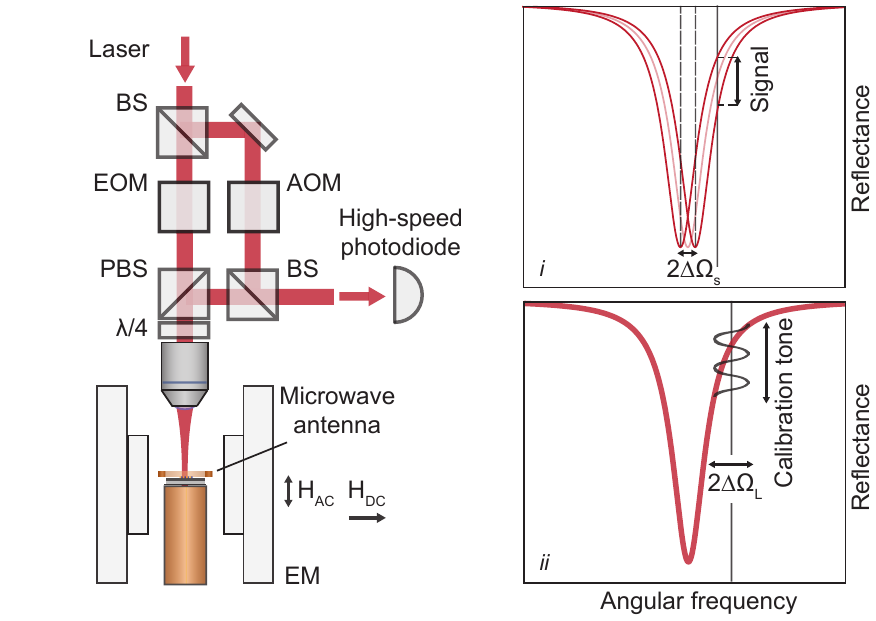}
\caption{\label{SIf:setup_calib} \textbf{Heterodyne scheme to calibrate the magnon-induced exciton shift $\boldsymbol{\Delta\Omega_s}$ and the coupling rate $g$.} 
A local oscillator is shifted in frequency by an acousto-optic modulator (AOM) and recombined with the signal reflected from the sample. The resultant signal is measured with a high-speed photodiode and analyzed on a spectrum analyzer (EM: electromagnet, PBS: polarizing beamsplitter, BS: beamsplitter, $\lambda/4$: quarter-wave plate).
We perform the measurement in two successive steps. First, we acquire the signals due to the modulation of the optical reflection induced by the magnons\textcolor{ForestGreen}{, see Eq.~(\ref{eq:modm})}, signing the shift of the exciton resonance frequency by $\pm\Delta \Omega_s/2\pi$ (i). Then, with no driven magnons and constituting the calibration tone, we measure the signals due to the modulation of the optical reflection induced by the modulation of the apparent laser frequency by $\Delta \Omega_L/2\pi$ produced by an electro-optic modulator (EOM)\textcolor{ForestGreen}{, see Eq.~(\ref{eq:modE})} (ii). 
The value $\Delta \Omega_L$ is calibrated at the beginning of any set of measurements (see Sec.~\ref{txt:calibEOM}). 
This calibration tone scales the magnon-induced exciton Zeeman shift $\Delta\Omega_s$ \textcolor{ForestGreen}{(see Sec.~\ref{subsec:deltaomegas})}. Determining the number of driven magnons $n_\mathrm{magnon}$ (see Sec.~\ref{txt:nmagnon}) then leads to the extraction of the magnon--exciton coupling rate $g$ as in Fig.~4.  
}
\end{figure}

\paragraph*{}
The calibration consists in comparing the optical reflection modulation induced by the magnons, with the optical reflection modulation induced by a known modulation of the laser frequency. The calibration setup is presented in Fig.~\ref{SIf:setup_calib}. We divide the original laser path into two: one illuminating the sample from which we get the reflection signal from, and the other, going through an acousto-optic modulator shifting its frequency by $\omega_\mathrm{A}/2\pi = 80$\,MHz, which constitutes a local oscillator. The optical signal reflected from the sample is retrieved with a circulator consisting of a polarizing beamsplitter and a quarter-wave plate, to be recombined with the local oscillator before reaching the high-speed photodiode. 
\paragraph*{}
The magnon-induced signals are obtained by driving the magnons in the YIG film at the FMR frequency with a given microwave drive power $P_\mathrm{MW}$. \textcolor{Blue}{As expressed in Eq.~(\ref{eq:omexmod}),} the driven magnons modulate the excitonic resonance frequency by $\Delta \Omega_s~=~g \sqrt{n_\mathrm{magnon}}$ (Fig.~\ref{SIf:setup_calib}i). The number of excited magnons $n_\mathrm{magnon}$\textcolor{Blue}{, given by Eq.~(\ref{eq:nmagnongen}),} is independently calibrated (see Sec.~\ref{txt:nmagnon}). 
\paragraph*{}
Switching off the magnon excitation, we create calibration tones using an electro-optic modulator (EOM). Driven at the FMR frequency, the EOM induces a phase modulation of the laser impinging on the sample. The induced laser maximum frequency deviation $\Delta \Omega_L/2\pi$, independently calibrated (see Sec.~\ref{txt:calibEOM}), is responsible for an amplitude modulation of the signal reflected from the sample (Fig.~\ref{SIf:setup_calib}ii), which is later recombined with the local oscillator. Assuming the nature of the addressed excitons to be the same as in the magnon-induced case, the comparison of the calibration tones and the magnon-induced signal grants access to the magnon-induced Zeeman shift $\Delta\Omega_s$ \textcolor{ForestGreen}{(see Sec.~\ref{subsec:deltaomegas})}. 
These measurements performed while ramping the number of driven magnons $n_\mathrm{magnon}$ leads to the evaluation of the magnon--exciton coupling rate $g$.
\subsection{Evaluation of the number of magnons $n_\mathrm{magnon}$}
\label{txt:nmagnon}
The number of magnons excited in the magnetic film is determined with Eq.~(\ref{eq:nmagnongen}) for a resonant excitation ($\omega_d = \omega_m$):
\begin{equation*}
n_\mathrm{magnon} = \dfrac{4 \gamma_e}{\gamma_m^2} \dfrac{P_\mathrm{MW}}{\hbar\omega_m}.
\end{equation*}
\color{Blue} The relevant parameters are evaluated by adjusting the FMR absorption signal $|S_{11}|$ with the model derived from the input--output theory~\cite{Clerk2010} in Eq.~({\ref{eq:S11}) with typically $\gamma_0/2\pi \sim 1.5$\,MHz and $\gamma_e/2\pi \sim 150\,$kHz.
An example of successive FMR measurements while ramping the driving power and the deduced number of magnons are shown in Fig.~\ref{SIf:nmag}.
\color{ForestGreen} Note that the magnons in the magnetostatic mode we are studying have long wavelength and thus delocalized over the entire sample. These delocalized magnons, whose number is defined and estimated for the whole sample, will interact with locally excited excitons.
\color{black}
\begin{figure}[b]
\includegraphics[scale=1]{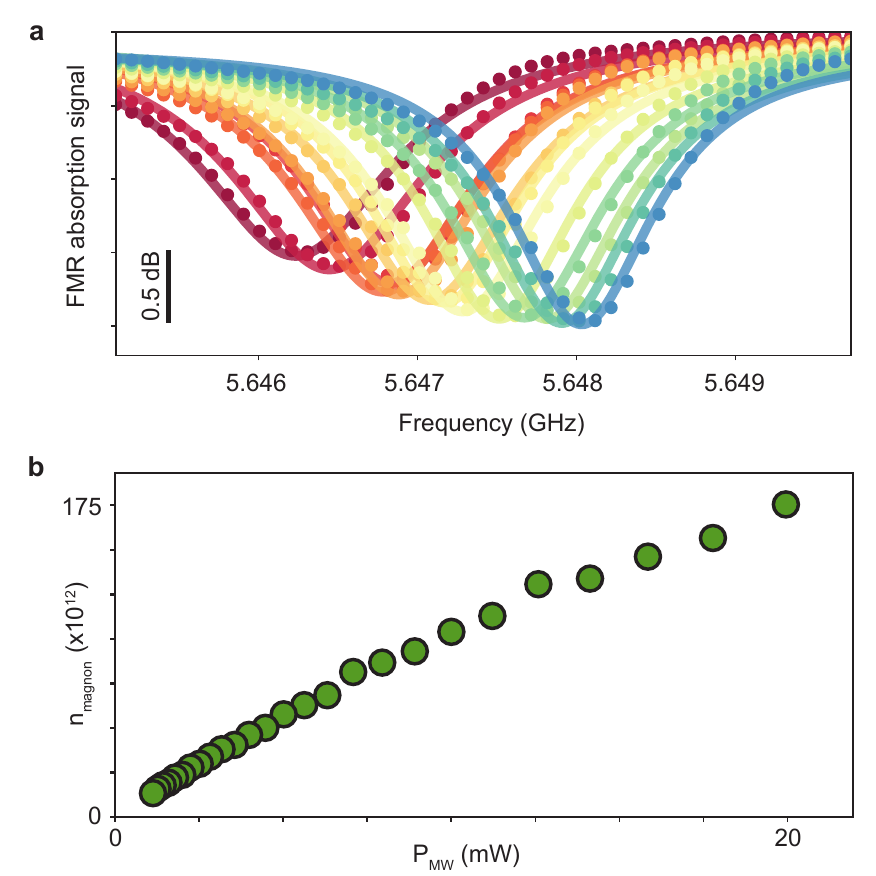}
\caption{\label{SIf:nmag}\textbf{Determination of the number of excited magnons.} \textbf{a,} FMR absorption signals for driving microwave power $P_\mathrm{MW}$ from 0.5\,dBm (blue) to 13\,dBm (red) fitted with Eq.~(\ref{eq:S11}). The slight deviations in the FMR frequency are most likely due to temperature variations. \textbf{b,} Number of magnons $n_\mathrm{magnon}$ in the uniform magnetostatic mode deduced from Eq.~(\ref{eq:nmagnongen}) with the analysis of the FMR absorption signals for a resonant excitation.}
\end{figure}

\subsection{EOM calibration}
\label{txt:calibEOM}
\paragraph*{}
The determination of the magnon--exciton coupling strength is based on the knowledge of a reference, the laser maximum frequency deviation $\Delta\Omega_L/2\pi$ induced by the passage of the beam through the EOM. The calibration of the value of $\Delta\Omega_L/2\pi$ is thus performed at the beginning of every measurement to take into account possible changes in the illumination or temperature, following the procedure described below.
Passing through the EOM, the laser beam is phase-modulated such that its electric field $E_E$ reads
\textcolor{ForestGreen}{
\begin{equation}
E_E = E_{i} \,\, e^{-i \Omega_L t} \,\, e^{-i \beta\sin\omega_E t} =  E_{i}\,\,e^{-i \Omega_L t} \sum_{m=-\infty}^{\infty} J_m (\beta) \, e^{-i m\omega_E t} 
\label{eq:EEOM}
\end{equation}}
and its instantaneous angular frequency is given by 
\begin{equation*}
\frac{\partial}{\partial t} \left(\Omega_L t + \beta\sin\omega_E t \right) =  \Omega_L + \beta \omega_E \cos\omega_E t =  \Omega_L + \Delta\Omega_L\cos\omega_E t
\end{equation*}
where $\omega_E/2\pi$ is the EOM drive frequency, $J_m$ is the Bessel functions of the first kind and $\beta$ is the modulation depth, whose determination leads to $\Delta\Omega_L = \beta \omega_E$. At low EOM driving power $P_E$, $\beta = \alpha \sqrt{P_E}$.

The beating of the phase-modulated laser with the local oscillator, whose frequency is shifted by $\omega_A/2\pi$, is detected with the high-speed photodiode and observed on the spectrum analyzer. The resulting power spectrum $C_E$ presents distinct signatures, in particular: 
\begin{align*}
C_E[\omega_A] &\propto J_0^2(\beta) = J_0^2(\alpha \sqrt{P_E}), \\ 
C_E[\omega_E\pm\omega_A] &  \propto J_1^2(\beta) = J_1^2(\alpha \sqrt{P_E}).
\end{align*}
\begin{figure}
\includegraphics[scale=1]{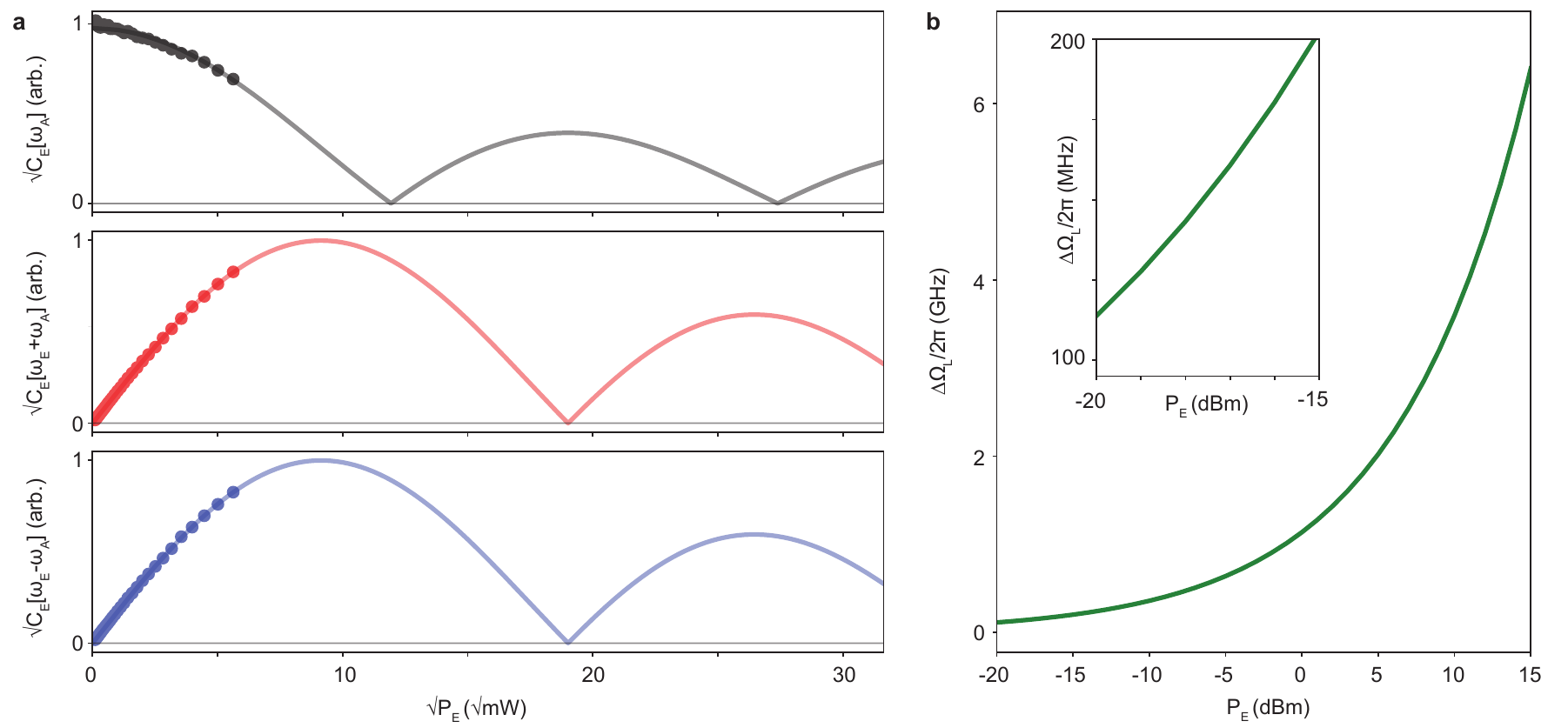}
\caption{\label{SIf:calibEOM}\textbf{Calibration of the electro-optic modulator.} \textbf{a,} Evolution of the signals $C_E$ at $\omega_A$ and $\omega_E\pm\omega_A$ with the driving EOM power $P_E$. The solid lines corresponding to fits proportional to $|J_0(\alpha\sqrt{P_E})|$ and $|J_1(\alpha\sqrt{P_E})|$ for $\omega_A$ and $\omega_E\pm\omega_A$, respectively, leading to $\alpha$.
\textbf{b,} Laser maximum frequency deviation $\Delta\Omega_L/2\pi$ induced by a given EOM microwave drive $P_E$ at $\omega_E/2\pi = 5.65\,$GHz from the ramp fits.
}
\end{figure}	
\paragraph*{}
The EOM drive frequency $\omega_E/2\pi$ is chosen equal to the FMR frequency $\omega_m/2\pi$, determined after FMR absorption measurements. The calibration of $\Delta\Omega_L(P_E) = \omega_E \, \alpha \sqrt{P_E}$ consists in ramping the microwave power $P_E$ driving the EOM and measuring the induced beating signals $C_E[\omega_A]$ and $C_E[\omega_E\pm\omega_A]$ to determine unambiguously $\alpha$, as illustrated in Figure~\ref{SIf:calibEOM}. 
Working at $P_E = -20$\,dBm at $\omega_E/2\pi$ = 5.65\,GHz leads to a maximum frequency deviation $\Delta\Omega_L/2\pi = 113.8\,$MHz.  

\color{ForestGreen}
\subsection{Evaluation of the exciton resonance shift from calibrated spectra}
\label{subsec:deltaomegas}
Here, we explain in details the evaluation of the experimental exciton resonance shift $\Delta \Omega_s$ from the measured spectra.
For the EOM phase-modulated case, the expression of the electric field given by Eq.~(\ref{eq:EEOM}) can actually be simplified as the modulation depth $\beta \sim 0.02 \ll 1$ to
\begin{equation}
E_E = E_{i}\,\, \left(e^{-i \, \Omega_L t} + \frac{\beta}{2} \, e^{-i \, (\Omega_L+\omega_E) t} - \frac{\beta}{2} \, e^{-i \, (\Omega_L-\omega_E) t}\right).
\end{equation}
We consider optical photons with angular frequencies  $\Omega_L$ and $\Omega_L\pm\omega_E$. \textcolor{Blue}{We omit here the subscript in $r_\mathrm{N_L}$ for sake of readability and write $r[\Omega]$ the reflection coefficient of the flake at the angular frequency $\Omega$.}
The electric field reflected from the TMD reads:
\begin{equation}
E_{rE} = \textcolor{Blue}{-r E_{E}}= \textcolor{Blue}{-}E_{i}\,\, \left(r[\Omega_L] \, e^{-i \, \Omega_L t} + r[\Omega_L+\omega_E] \frac{\beta}{2} \, e^{-i \, (\Omega_L+\omega_E) t} - r[\Omega_L-\omega_E]\frac{\beta}{2} \, e^{-i \, (\Omega_L-\omega_E) t}\right). \label{eq:modE}
\end{equation}

Assuming that the TMD reflectivity spectrum is linear on the range $\Omega_L \pm \omega_E$, we can write
\begin{equation}
\label{eq:der}
r[\Omega_L\pm\omega_E] = r[\Omega_L] \pm \frac{\partial r}{\partial \omega} \omega_E \mathrm{,}
\end{equation}
with $\partial r/\partial \omega \equiv \partial r_\mathrm{N_L}/\partial(\Omega_\mathrm{ex}-\Omega_L)$. The form of $\partial r/\partial \omega$ is explicitly described in the monolayer case in Eq.~(\ref{deltarwm}).

The optical signal coming from the sample is beaten with the local oscillator with an electric field
\begin{equation}
E_{A} = E_{Ai} \,\, e^{-i (\Omega_L+\omega_A) t}.
\end{equation}
Close to the optical resonance ($\Omega_L \sim \Omega_\mathrm{ex}$), the detected intensity $I_E$ in the EOM case follows
\begin{align*}
I_E \propto |E_{A} + E_{rE}|^2 = \textcolor{Blue}{\bigg\lvert-} 2 \, E_{Ai} \, E_{i} \, r[\Omega_L]& \,\,\,\, \cos \omega_A t  \\
+ 2 E_{i}^2 \, r[\Omega_L] \, \frac{\partial r}{\partial \omega} \Delta\Omega_L& \,\,\,\,  \cos \omega_E t  \\
\textcolor{Blue}{-} E_{Ai} \, E_{i} \,\, \left(\beta r[\Omega_L] + \frac{\partial r}{\partial \omega} \Delta \Omega_L \right)& \,\,\,\, \cos (\omega_E-\omega_A)t  \\
\textcolor{Blue}{+} E_{Ai} \, E_{i} \,\, \left(\beta r[\Omega_L] - \frac{\partial r}{\partial \omega} \Delta \Omega_L \right)& \,\,\,\, \cos (\omega_E+\omega_A)t  \\
- \frac{E_i^2}{2} \left(\beta^2 r[\Omega_L]^2-\left(\frac{\partial r}{\partial \omega}\right)^2 \Delta\Omega_L^2\right)& \,\,\,\, \cos 2\omega_E t \\
+ E_{Ai}^2 + \frac{E_{i}^2}{2} \left((2+\beta^2) r[\Omega_L]^2+ \left(\frac{\partial r}{\partial \omega}\right)^2 \Delta\Omega_L^2\right)\bigg\rvert& \mathrm{.}
\end{align*}

In the driven magnon case, the Zeeman effect induces an amplitude modulation such that the electric field reads
\begin{equation}
E_{r} =  \textcolor{Blue}{- [r + \delta r(t)] E_{i} e^{-i \Omega_L t}} = \textcolor{Blue}{-}E_{i} \,\, \left(r[\Omega_L] e^{-i \Omega_L t} + \frac{\partial r}{\partial \omega} \frac{\Delta\Omega_s}{2} \,  e^{-i \, (\Omega_L+\omega_m) t}
 + \frac{\partial r}{\partial \omega} \frac{\Delta\Omega_s}{2} \, e^{-i \, (\Omega_L-\omega_m) t}\right). \label{eq:modm}
\end{equation}
This signal is beaten with the local oscillator such that the detected intensity $I_m$ follows
\begin{align*}
I_m \propto |E_{A} + E_{r}|^2 = \bigg\lvert\textcolor{Blue}{-}2 \, E_{Ai} \, E_{i} \, r[\Omega_L]& \,\,\,\, \cos \omega_A t\\
+ 2 E_{i}^2 \, r[\Omega_L] \, \frac{\partial r}{\partial \omega}\Delta \Omega_s& \,\,\,\, \cos \omega_m t \\
\textcolor{Blue}{-} E_{Ai} \, E_{i} \,\, \frac{\partial r}{\partial \omega} \Delta \Omega_s& \,\,\,\, \cos (\omega_m-\omega_A)t \\
\textcolor{Blue}{-} E_{Ai} \, E_{i} \,\, \frac{\partial r}{\partial \omega} \Delta \Omega_s& \,\,\,\, \cos (\omega_m+\omega_A)t \\
+ \frac{E_{i}^2}{2} \, \left(\frac{\partial r}{\partial \omega}\right)^2 \Delta \Omega_s^2& \,\,\,\, \cos 2\omega_m t \\
+ E_{Ai}^2 + E_{i}^2 \left(r[\Omega_L]^2 + \frac{1}{2} \, \left(\frac{\partial r}{\partial \omega}\right)^2 \Delta \Omega_s^2\right)\bigg\rvert&  \mathrm{.}
\end{align*}
\paragraph*{}
On the spectrum analyzer, we acquire the power spectral density of the photodiode output for the EOM and the magnon-driven case, $S_E[\omega] \propto I_E[\omega]^2$ and $S_m[\omega] \propto I_m[\omega]^2$ respectively, at the carrier ($\omega_E = \omega_m$) and sidebands ($\omega_m \pm \omega_A$) frequencies. The information on $\Delta \Omega_s$ can be redundantly extracted on the carrier and on the sidebands. 
\paragraph*{}
From the carrier, the transformation is straightforward
\begin{equation}
\label{eq:viacarrier}
\Delta \Omega_s = \frac{S_m[\omega_m]^\frac{1}{2}}{S_E[\omega_m]^\frac{1}{2}} \, \Delta \Omega_L.
\end{equation}
From the sidebands, 
assuming the electric transfer function of the measurement chain to be flat on $\omega_m \pm \omega_A$,
\begin{equation}
\label{eq:viasideband}
\Delta \Omega_s = 2\,\frac{S_m[\omega_m\pm\omega_A]^\frac{1}{2}}{S_E[\omega_m-\omega_A]^\frac{1}{2} - S_E[\omega_m+\omega_A]^\frac{1}{2}} \, \Delta \Omega_L  \mathrm{.}
\end{equation}
The evaluation using the sidebands and Eq.~(\ref{eq:viasideband}) being based on the difference between two similar signals could raise concerns as it may be within the detection precision of the spectrum analyzer. 
While actually leading to analogous results with both methods (from sidebands: $\hbar g_\pm = 1.8 - 4.9$\,feV, from carrier [as in the main text]: $\hbar g_\pm = 3.1 - 4.4$\,feV), we show the results from the carrier using Eq.~(\ref{eq:viacarrier}) as the analysis is more straightforward. 

\color{ForestGreen}
\section{Discussions on the estimate of the magnon--exciton coupling strength}
\color{black}
\subsection{Estimation of the dipolar magnetic field induced by the magnons}
\label{txt:dipolar}
The theoretical evaluation of the magnon--exciton coupling rate $g$ is not trivial when it originates from an interfacial exchange effect since it involves calculations beyond the well-established density functional theory. Here, we provide a theoretical estimate of $g$ assuming that the coupling is purely from the dipolar field created by the magnons \textcolor{ForestGreen}{as a benchmark}.  \textcolor{Blue}{The obtained value is then compared with experimental results in Sec.~\ref{ss:currentg}.}

\subsubsection*{Dipolar field from a static out-of-plane magnetization}
Before delving into the calculation of the dipolar field produced by the magnons, let us consider a simpler case. Suppose that the film is saturated along the $z$-axis, normal to its plane. We calculate the dipolar field $B_z$ at the film surface created by the magnetic moment $\mathcal{M}_z = M_{s}V$, with $V= w^2 \times d$ the film volume. 
If the lateral dimensions of the magnetic film were infinitely large, the dipolar magnetic field $B_z$ would cancel out. However for a finite-size sample, $B_z$ is small but non-zero. It can be considered as emerging from the edge current~\cite{Purcell1985}
\begin{equation}
I = \frac{\mathcal{M}_z}{w^2},
\end{equation}
which is flowing in a rectangular ribbon of width $d$ enclosing the area $w^2$. From the Biot-Savart law, the out-of-plane dipolar magnetic field at the center of the film is
\begin{equation}
B_{z} = \mu_{0} \frac{2\sqrt{2}}{\pi}  \frac{d}{w} M_{s}. \label{eq:Bz}
\end{equation}

\subsubsection*{Dipolar field from an oscillating out-of-plane magnetization}

We suppose now that the magnetization is saturated along the $y$-axis. Nominally the static out-of-plane field is absent, but when magnons are driven, there is an oscillating transverse magnetic field with a non-zero out-of-plane component $B_{z}$. 
The transverse field is produced by the magnetic moment 
\begin{align}
\mathcal{M}_{z} = m_{z}V  
						= \left(\sqrt{\frac{H_\mathrm{DC}}{2H_\mathrm{DC}+M_{s}}} \sqrt{2}\, \frac{M_{s}}{\sqrt{N}} {\sqrt{n_\mathrm{magnon}} \cos \omega_{d}t}\right)V, \label{eq:Bz2}
\end{align}
\textcolor{Blue}{where the magnetization $m_{z}$ is given by Eq.~(\ref{eq:mz2}).}
We can repeat the argument which leads to the magnetic field $B_z$ in Eq.~(\ref{eq:Bz}) but with this magnetic moment \textcolor{Blue}{to obtain}
\begin{equation}
B_{n_\mathrm{magnon}} = \mu_0 \frac{2\sqrt{2}}{\pi} \frac{d}{w} \left( \sqrt{\frac{H_\mathrm{DC}}{2H_\mathrm{DC}+M_{s}}} \sqrt{2}\, \frac{M_{s}}{\sqrt{N}}\sqrt{n_\mathrm{magnon}} \cos \omega_{d}t \right). \label{eq:Bn}
\end{equation}
The amplitude of the oscillating dipolar magnetic field produced by a single magnon ($n_\mathrm{magnon} = 1$) is then given by:
\begin{equation}
B_{D} = \mu_0\frac{4}{\pi} \frac{\sqrt{d}}{w^2} \,\, \sqrt{\frac{H_\mathrm{DC}}{2H_\mathrm{DC}+M_s}}\frac{M_s}{\sqrt{n}},
\end{equation}
with $n = N/V = N/w^2d$ the spin density of the magnetic substrate (in YIG~\cite{Stancil2009}, $n= 2.1\times10^{22}$\,cm$^{-3}$). In our experimental conditions, it results in $B_{D} = 0.1\,$pT. From there, we can deduce the magnon--exciton coupling strength originating purely from this dipolar field $\hbar g_D \sim 1.2 \times 10^{-17}\,$eV \textcolor{ForestGreen}{[$g_D/2\pi \sim 3\,$mHz]}. 

\color{ForestGreen}
\subsection{Current experimental estimate and prospectives of the magnon-exciton coupling strength}
\label{ss:currentg}
\subsubsection*{Measured magnon-exciton coupling strength and origin of the observed effects}
\paragraph*{} 
The measured magnon-exciton coupling strength 
\begin{equation}
\hbar g_{\pm} \sim 4~\mathrm{feV} \quad [g_\pm/2\pi \sim 1\,\mathrm{Hz}]
\end{equation}
reported in the main text is more than two orders of magnitude higher than the value expected from a pure dipolar origin. \textcolor{Blue}{With the observation of valley-resolved features of the Zeeman effects on multilayer flakes (Fig.~2) and the strong thickness dependence of $\delta r_\mathrm{N_L}$ (Fig.~3 and Sec.~\ref{subsec:drfNL}), the measured value of the coupling strength supports furthermore that exchange interactions are at play.}
\subsubsection*{Interface quality}
In order to evaluate the quality of the interface of our sample, we can compare the proximity effects observed in our experiment with values previously reported.
The single-magnon Zeeman shift can be considered as stemming from the quantum fluctuation of the mean magnetization of the magnetic film, scaling as $M_s/\sqrt{N}$, with, as a reminder, $N \sim 5 \times 10^{18}$ the total number of spins in the YIG magnetic film and $M_s$ its saturation magnetization, lying within the film plane in our experiment. 

This value can then be translated into some equivalent static Zeeman shift induced by an out-of-plane mean magnetization. The involved magnetization would be $\sqrt{N}$ larger, leading to an equivalent Zeeman shift of
\begin{equation}
\hbar g_{\pm} \sqrt{N} \sim 10\,\mathrm{\mu eV} \mathrm{.} 
\end{equation}
Keeping in mind that the materials and thus the interface are intrinsically different, this value can be examined in the light of previously reported Zeeman shifts through static magnetic proximity effects, such as 1\,meV for WSe$_{2}$ on EuS and 10\,meV for WS$_{2}$ on EuS~\cite{Zhao2017,Norden2019}.

Provided that saturation magnetization in EuS is about one order of magnitude higher than YIG~\cite{Zhao2017, Stancil2009}, our estimate for the coupling YIG/MoSe$_{2}$ is then only about one-to-two orders of magnitude lower than the measurements done on the interfaces of WSe$_{2}$/EuS and WS$_{2}$/EuS. 
TEM cross-sectional measurements suggest that the interface could be improved by guaranteeing a better homogeneity and flatness of the YIG surface.
Improving the interface quality toward the values reported in the literature~\cite{Zhao2017,Norden2019} and beyond (100\,meV for MoTe$_{2}$/EuO was suggested by theoretical calculations~\cite{Qi2015}) would lead to stronger magnon-exciton coupling.

\subsubsection*{Reducing the magnetic substrate volume}
Once we reduce the volume of magnetic substrate and thus confine magnons within a smaller volume, the magnon-exciton coupling would increase significantly. 
Considering a thinner magnetic film with a lateral size comparable to the TMD flake, with a volume $V_r = 100\,\mathrm{nm} \times 5\,\mathrm{\mu m}\,\times 5\,\mathrm{\mu m}$, the total number of spin would then be 8 orders of magnitude smaller than in the current film, that is
\begin{equation}
N_{r} = n \times V_{r} \sim 5 \times 10^{10}. \label{eq:N1}
\end{equation}
The single-magnon Zeeman shift being proportional to the inverse of the square-root of the total number of spins, it will be improved from the current value by $\sqrt{\frac{N}{N_{r}}}$ to 
\begin{equation}
\hbar g_{r} 
\sim 40\,\mathrm{p eV} \quad [g_{r}/2\pi \sim 10\,\mathrm{kHz}]. \label{eq:g1}
\end{equation}
}

\section{Additional experimental data}
\subsection{Photoluminescence measurements}
\begin{figure}
\includegraphics[scale=1]{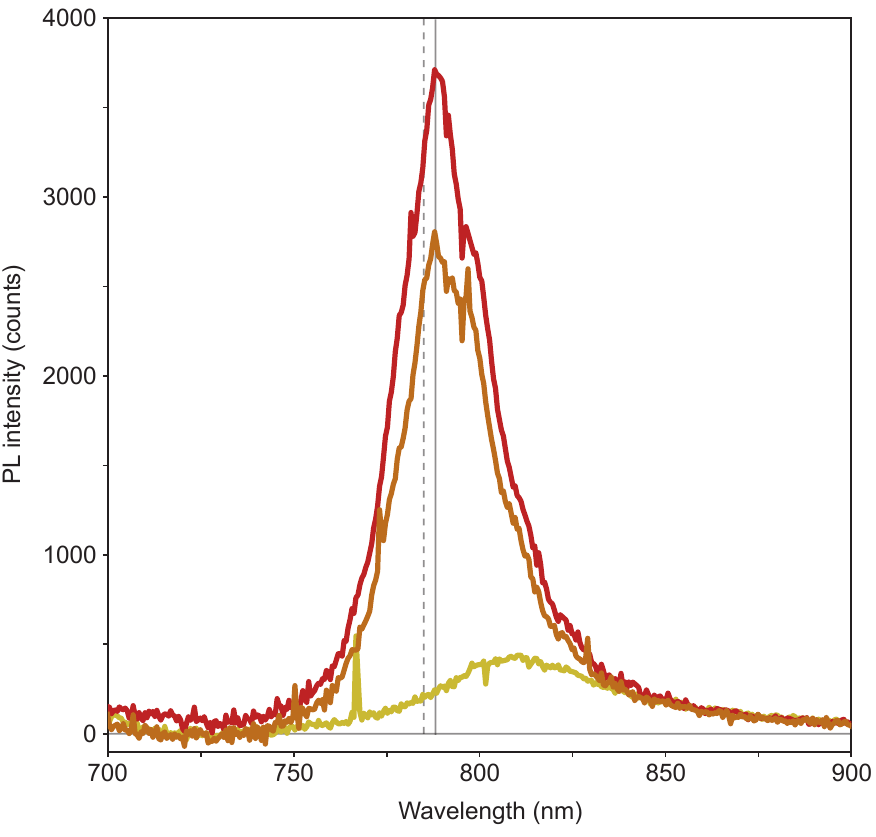}
\caption{\label{SIf:PL}Photoluminescence (PL) measurements performed with a laser at 633\,nm on different MoSe$_2$ flakes on YIG/GGG, identified as monolayers in red and orange and as bilayer in yellow. The solid and dashed vertical lines indicate respectively 788\,nm, the center wavelength of the PL spectra at 300\,K for MoSe$_2$ monolayers, and $\lambda_L = 785$\,nm the wavelength of the laser used for the reflection measurements in the main text.}
\end{figure}

As shown in Fig.~\ref{SIf:PL}, we performed photoluminescence measurements on the MoSe$_2$ flakes after their transfer on the YIG substrate, under no external magnetic field ($H_\mathrm{DC}=0$). The exciton resonance wavelength of monolayers is 788\,nm at 300\,K as for typical MoSe$_2$ monolayers on Si/SiO$_2$~\cite{Arora2015}.
\color{ForestGreen}
Note that since we are working at room temperature, the exciton linewidth ($50-100$\,meV) is large enough to be probed by a single wavelength, even if the exciton resonance frequency slightly varies with the thickness~\cite{Arora2015}. 
\color{black}

\subsection{Ferromagnetic resonance frequency}
\begin{figure}
\includegraphics[scale=1]{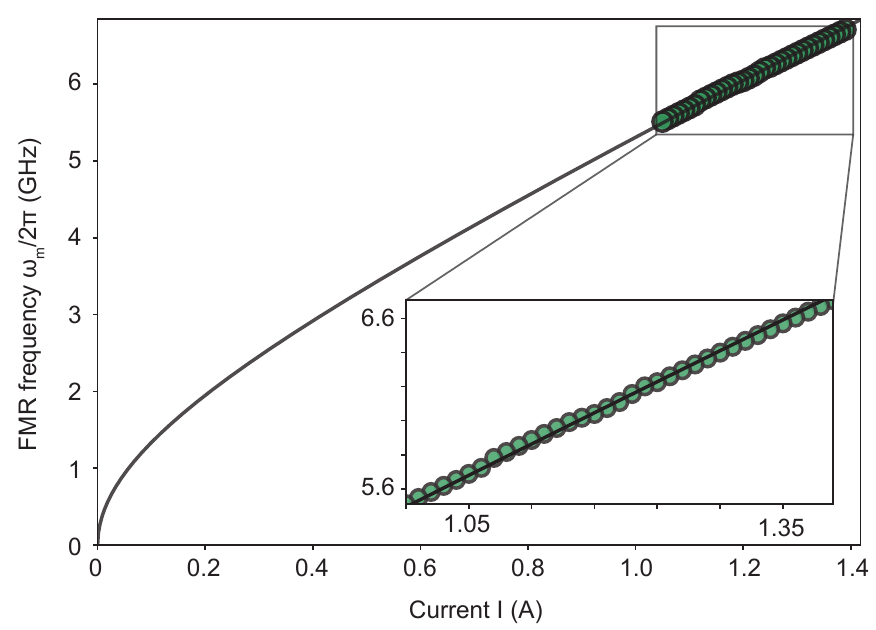}
\caption{\label{SIf:wm} Evolution of the measured FMR frequency with the applied current flowing in the electromagnet. The black line follows Eq.~(\ref{eq:freqmagnon}) with $M_s = 140$\,kA/m, $\gamma/2\pi = 28$\,GHz/T and $\alpha_H = 0.12$\,T/A.
}
\end{figure}

The angular frequency of the ferromagnetic resonance (FMR) for the uniformly precessing mode in the magnetic film can be tuned by the tangential static field $H_\mathrm{DC}$ as expressed in Eq.~(\ref{eq:freqmagnon}):
\begin{equation*}
\omega_m = \mu_0 \gamma \sqrt{H_\mathrm{DC}\left(H_\mathrm{DC}+M_s\right)},
\end{equation*} 
where $\gamma$ is the gyromagnetic ratio and $M_s$ is the saturation magnetization of the film (typically for YIG~\cite{Stancil2009}: $\gamma/2\pi = 28$\,GHz/T and $M_s = 140$\, kA/m). The static magnetic field depends linearly on the current $I$ flowing in the electromagnet, such that $\mu_0 H_\mathrm{DC} = \alpha_H I$, with $\alpha_H = 0.12$\,T/A.
Figure~\ref{SIf:wm} presents the experimental dependence of the FMR frequency with the current $I$, following Eq.~(\ref{eq:freqmagnon}). 

\subsection{Notes on experimental signatures of multilayer flakes}

\textcolor{ForestGreen}{
From static measurements of the reflectance across the sample with flake layers known from atomic force microscopy, we deduce $r_\mathrm{1L} \sim 6\%$ from the comparison of the static reflection coefficient $r_\mathrm{N_L}$ with the modeled values given by Eq.~(\ref{eq:r}) and presented in Fig.~\ref{SIf:reflpart2}.
\begin{figure}[h]
\includegraphics[scale=1.1]{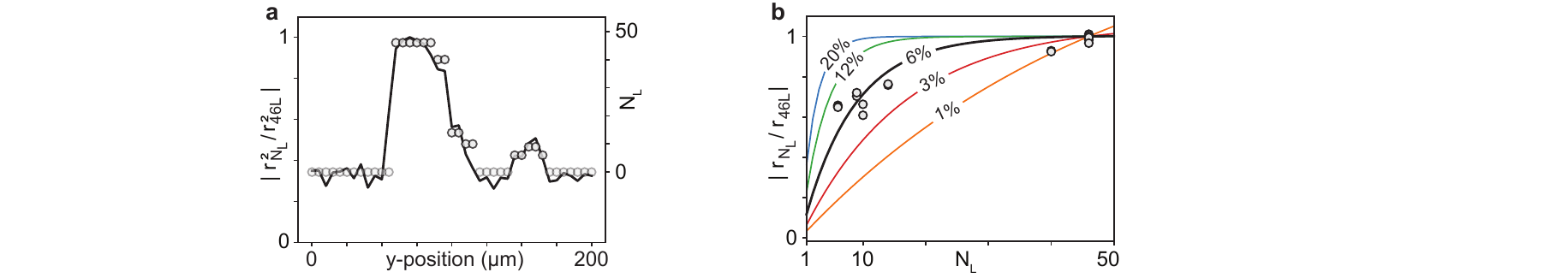}
\caption{\label{SIf:reflpart2} \textcolor{Blue} {
\textbf{Experimental determination of the effective monolayer reflection coefficient.} \textbf{a,} Normalized static reflectance $|r_\mathrm{N_L}^2/r_\mathrm{46L}^2|$ measured across a section of the sample (black solid line), moving the laser spot along the $y$-axis similarly to Fig.~3\textbf{a}(i), superimposed with the number of layers deduced from atomic force microscopy measurements (grey circles).}
 \textbf{b,} Reported reflection coefficients as a function of the number of layers with thick layers as a reference ($N_L = 46$) to determine the effective monolayer reflection coefficient $r_\mathrm{1L} = 6\%$. The solid lines correspond to the model in Eq.~(\ref{eq:r}) for various $r_\mathrm{1L}$: 20\%, 12\%, 6\%, 3\% and 1\% in blue, green, black, red and orange, respectively.}
\end{figure}
}

\color{Sepia}
\subsubsection*{Dependence of the apparent dynamical perturbation of the optical reflection on the number of layers}
\label{subsec:drfNL} 
We use the model we presented in Sec.~\ref{sec:genmultiflake} to access the apparent dynamical perturbation $\delta r_\mathrm{N_L}$ independently from the enhanced static reflection coefficient $r_\mathrm{N_L}$. 
In the main text, we consider $\Delta R[\omega] = |R_+[\omega] - R_-[\omega]|$, with $R_\pm[\omega] \propto E_r[\omega]^2$ the reflected optical signal originating from $\sigma_\pm$ circular light polarizations. From Eq.~(\ref{eq:Erwom}), assuming equal static coefficients and opposite dynamical perturbations for $\sigma_\pm$ polarization as observed in Fig.~1, then
\begin{equation}
\frac{\Delta R[\omega_m]}{r_{N_L}} \propto |\delta r_{N_L}[\omega_m]|\mathrm{.}
\end{equation} 
We fit the data representing $\Delta R[\omega_m]/r_{N_L}$ in Fig.~3, with a function proportional to  $\rho_\mathrm{N_L} + \rho_\mathrm{LR}$, where $\rho_{N_L} $ is the relative portion of light coming from the very bottom layer defined in Eq.~(\ref{SIeq:rhob}), and $\rho_\mathrm{LR}$ is a small constant modeling the long-range interaction and the instrumental detection background. This suggests that the apparent dynamical perturbation decreases as the number of layer increases in the same fashion as the portion of light coming back from the very bottom layer. In other words, the modulation of the light mostly originates from the closest layers to the magnetic substrate, suggesting a dominant short-range interaction.  The detection background, attributed to electronics, can be evaluated by averaging the whole spectra on bare YIG, accounting for $\sim8\%$ of the maximum value of the dataset.

\begin{figure}[h]
\includegraphics[scale=1]{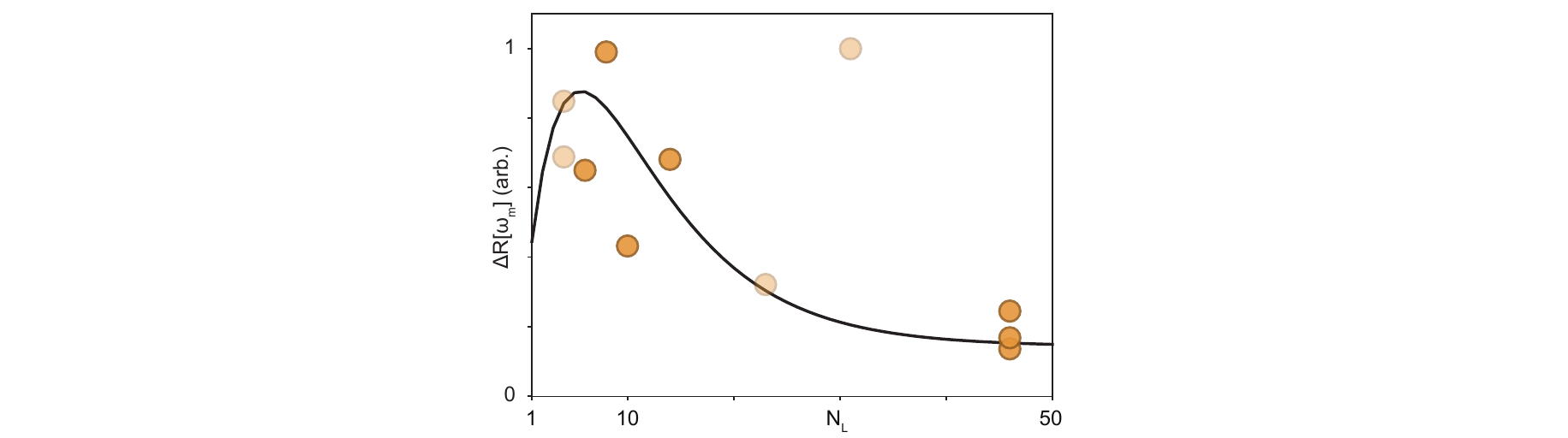}
\caption{\label{SIf:reflpartdynB} \textcolor{Sepia} {Differential optical reflectance $\Delta R$ at the FMR frequency as a function of the number of layers. Compared with the monotonically declining behaviour shown in Fig.~3\textbf{d} where $\Delta R[\omega_m]/r_\mathrm{N_L}$ is plotted, there is a maximum. 
The solid black line is a function proportional to $r_\mathrm{N_L} (\rho_\mathrm{N_L}+ \rho_\mathrm{LR})$, using the same fit parameters as in Fig.~3\textbf{d}, presenting an optimum at $N_L \sim 6$ that maximizes the reflected signal at the FMR frequency, as previously predicted in Fig.~\ref{SIf:reflpartdynA}.}} 
\end{figure}

As predicted in Sec.~\ref{sec:genmultiflake} and illustrated in Fig.~\ref{SIf:reflpartdynA}, the intensity of the reflected optical beam at the FMR frequency is the result of a competition between the static reflection coefficient $r_\mathrm{N_L}$ and the portion of light coming back from the bottom layer $\rho_\mathrm{N_L}$, which has a maximum for an optimum number of layers at a given $r_\mathrm{1L}$. In our experimental conditions with $r_\mathrm{1L} = 6\%$, the optimum number of layers is expected to be $\mathrm{N_L} = 6$. Figure~\ref{SIf:reflpartdynB} shows the differential optical reflectance $\Delta R$ at the FMR frequency as a function of the number of layers, indicating the maximum around $N_L \sim 6$. The agreement in Fig.~\ref{SIf:reflpartdynB} and Fig.~3\textbf{d} in the main text with the model described by Eq.~(\ref{eq:r}-\ref{SIeq:rhob}) and Eq.~(\ref{eq:Erwom}) also indicates that the decay of the reflection from the bottom layer is dominated by the reflection loss rather than the absorption. 

\color{black}

\subsubsection*{Optical response of a trilayer MoSe$_2$ flake on YIG}
\begin{figure}[h!]
\includegraphics[scale=1]{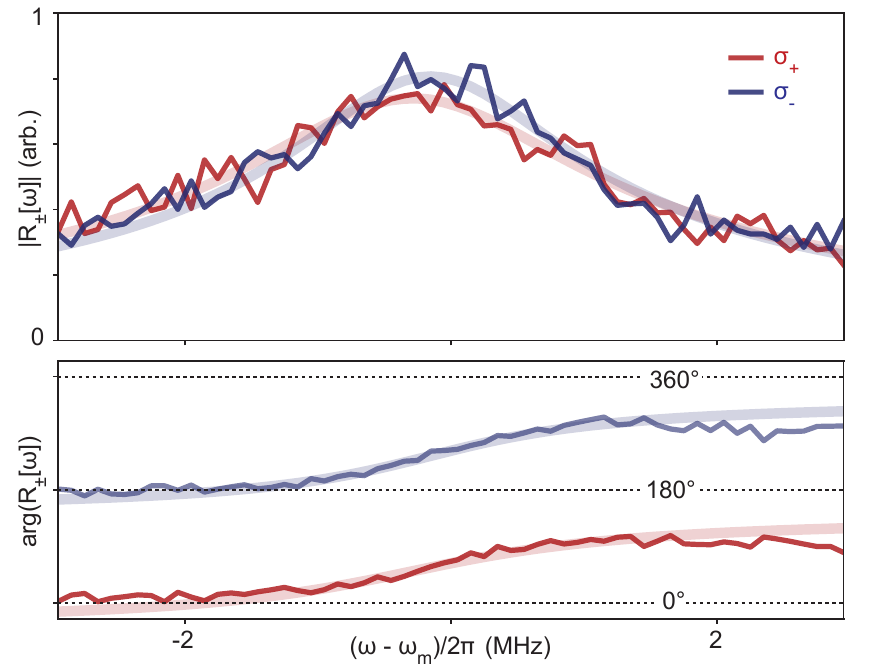}
\caption{\label{SIf:3L} Optically-probed FMR spectra $R_+[\omega]$ and $R_-[\omega]$ from a trilayer MoSe$_2$ flake on YIG (in red and blue, respectively). The translucent lines correspond to Lorentzian fits.}
\end{figure}

We perform similar measurements as presented in Fig.~2 on trilayer MoSe$_2$ flakes, which offer a tenuous signal-to-noise ratio due to their low reflectance. As shown in Fig.~\ref{SIf:3L}, the main features presented in the main text are preserved with a trilayer flake. 
\pagebreak
\color{black}
\bibliography{biblio_formate}